\documentclass[format=acmsmall, review=false, screen=true]{acmart}

\usepackage{booktabs} % For formal tables

\usepackage{makecell}
\usepackage{array}
\usepackage{varwidth}

\usepackage[ruled]{algorithm2e} % For algorithms

\SetAlFnt{\small}
\SetAlCapFnt{\small}
\SetAlCapNameFnt{\small}
\SetAlCapHSkip{0pt}
\IncMargin{-\parindent}

\usepackage{cleveref}
\usepackage[utf8]{inputenc}
\crefname{section}{§}{§§}
\Crefname{section}{§}{§§}
\usepackage[10pt]{moresize}
\usepackage[font={sf,scriptsize}]{caption}
\usepackage[font={sf,scriptsize}]{subcaption}
\usepackage{amsmath,amssymb,amsfonts}
\usepackage{mathtools,mathrsfs}
\usepackage{multirow}
\usepackage{rotating}
\usepackage{enumitem}
\newcommand{\maciej}[1]{\textcolor{blue}{[Maciej: #1]}}
\newcommand{\macb}[1]{\textbf{\textsf{#1}}}
\usepackage{scalerel,stackengine}
\stackMath
\newcommand\rwh[1]{%
\savestack{\tmpbox}{\stretchto{%
  \scaleto{%
      \scalerel*[\widthof{\ensuremath{#1}}]{\kern-.6pt\bigwedge\kern-.6pt}%
          {\rule[-\textheight/2]{1ex}{\textheight}}%WIDTH-LIMITED BIG WEDGE
            }{\textheight}% 
}{0.5ex}}%
\stackon[1pt]{#1}{\tmpbox}%
}

\renewcommand{\maciej}[1]{}

\usepackage{graphicx}

\usepackage{tikz}
\usetikzlibrary{tikzmark}

\usepackage{xpatch}
\expandafter\xpatchcmd
   \csname pgfk@/tikz/every picture/.@cmd\endcsname
            {\thepage}{\arabic{page}}{}{}

            \tikzstyle{comment} = [draw, fill=blue!70, text=white, text width=3cm, minimum height=1cm, rounded corners, align=left, font=\scriptsize]
            \tikzstyle{background_alg} = [draw, fill=blue!20, opacity=0.4, inner sep=4pt, rounded corners=2pt]

            \usetikzlibrary{shapes}
            \usetikzlibrary{plotmarks}
            \usetikzlibrary{calc,fit}

\acmDOI{0000001.0000001}

\begin{document}
% Title portion. Note the short title for running heads 

\newcolumntype{M}{>{\begin{varwidth}{8cm}}l<{\end{varwidth}}} %M is for Maximal column

%\title[Survey and Taxonomy of Graph Compression]{Survey and Taxonomy of Lossless and Lossy Graph Compression}
\title[Survey and Taxonomy of Lossless Graph Compression]{Survey and Taxonomy of Lossless Graph Compression\\and Space-Efficient Graph Representations}
%\title[Survey and Taxonomy of Lossless Graph Compression]{Survey and Taxonomy of Lossless Graph Compression}
\subtitle{Towards Understanding of Modern Graph Processing, Storage, and Analytics}

\author{Maciej Besta}
\affiliation{
  \institution{Department of Computer Science, ETH Zurich}
  \country{Switzerland}
}

\author{Torsten Hoefler}
\affiliation{
  \institution{Department of Computer Science, ETH Zurich}
  \country{Switzerland}
}

\begin{abstract}
Various graphs such as web or social networks may contain up to trillions of
edges. Compressing such datasets can accelerate graph processing by reducing 
the amount of I/O accesses and the pressure on the memory subsystem. Yet,
selecting a proper compression method is challenging as there exist a plethora
of techniques, algorithms, domains, and approaches in compressing graphs.
To facilitate this, we present a survey and taxonomy on \emph{lossless} graph
compression that is the first, to the best of our knowledge, to
\emph{exhaustively} analyze this domain. Moreover, 
our survey does not only categorize existing schemes, but also explains
\emph{key ideas}, discusses formal underpinning in selected works,
and describes the space of the existing compression schemes using three
dimensions: \emph{areas of research} (e.g., compressing web graphs),
\emph{techniques} (e.g., gap encoding),
and \emph{features} (e.g., whether or
not a given scheme targets dynamic graphs).
Our survey can be used as a guide to select
the best lossless compression scheme in a given setting. 
% 
% but also as a
% reference to \emph{understand} key ideas and insights.
% 
% Finally, our surveying methodology based on a
% formulation of a space of available designs can be used to enhance other
% similar works. 
% 
% Our evaluation illustrates that Log(Graph) achieves simplicity and tunable
% performance as well as compression ratios.
%
% Log(Graph) can be used to improve the design of any graph processing
% framework, algorithm implementation, or library on both fat shared-memory
% NUMA nodes and distributed-memory data centers and supercomputers.
%
\end{abstract}



%\thanks{\maciej{TODO}}

\maketitle

% The default list of authors is too long for headers}
\renewcommand{\shortauthors}{Maciej Besta and Torsten Hoefler}

\section{Introduction}

%\enlargethispage{\baselineskip}

Big graphs form the basis of many problems in machine learning, social network
analysis, and various computational
sciences~\cite{DBLP:journals/ppl/LumsdaineGHB07}.  Storage-efficient processing
of such graphs is becoming increasingly important for HPC and Big Data. First,
it may eliminate expensive I/O accesses. Moreover, it enables storing a larger
fraction of data in caches, potentially increasing performance. Next, it could
eliminate inter-node communication as graphs may fit in the memory of one node.
Finally, reducing required amounts of memory lowers costs of necessary
hardware.

There exists a plethora of graph compression schemes.
They cover various fields such as web graphs~\cite{boldi2004webgraph}, biology
networks~\cite{hayashida2010comparing, conway2011succinct}, or social
graphs~\cite{chierichetti2009compressing}.
Moreover, they follow different methodologies, for example
attempting to build a graph representation that asymptotically matches the
storage lower bound (so called \emph{succinct}
representations~\cite{Farzan:2013:SEA:2562345.2562632__}) or permuting vertex
integer labels to minimize the sum of differences between consecutive neighbors
of each vertex and encoding these minimized differences with variable-length
coding~\cite{boldi2009permuting_social}.
Next, there are different compression techniques that can be used in the context of 
any methodologies, for example reference
encoding~\cite{boldi2004webgraph}, Huffman degree
encoding~\cite{suel2001compressing}, and many others.
%
% Moreover, many of these techniques are lossless~\cite{suel2001compressing}
% while others are lossy~\cite{spielman2011graph}.
%
Finally, compression can be general and target any
graph~\cite{Farzan:2013:SEA:2562345.2562632__} or may be designed for a
particular class of graphs~\cite{blelloch2010succinct}.
%
% The points above are far from being complete and there are many others
% categories that one can use for analyzing the rich world of graph
% compression.

This paper aims to provide the first taxonomy and survey that attempts to cover
all the associated areas of lossless graph compression.  Our goal is to (1)
\emph{exhaustively} describe related work, (2) illustrate and explain the \emph{key
ideas} and the theoretical underpinning, and (3) \emph{systematically} categorize
existing algorithms, schemes, techniques, methodologies, and concepts.

\macb{What Is The Scope of This Survey?}
We focus on \emph{lossless} approaches and leave all the \emph{lossy} schemes
for future work. The main reason for this is the fact that the scope of
lossless graph compression is on its own very extensive, covering almost 
450 papers and various approaches, numerous techniques, countless algorithms,
and a large body of applications.

% \macb{Why The Name ``Condensing Graphs''?}
% %
% We decided to use the term ``condensing'' to refer in general to \emph{any}
% technique that aims to reduce the size of a graph. The reason behind this name
% is the fact that the existing literature does not agree on the generic term
% that covers all such schemes. Some authors~\cite{maneth2015survey} refer to
% compression as the generic name for reducing the size of graphs, while
% others~\cite{x} treat compression and succinctness as two different approaches
% to reduce graph sizes.  Then, there are other categories of research studies,
% such as implicit (or efficient) graph representations that can be viewed as the
% attempts to reduce the size of graphs but also fail to be included in either
% compression as seen by most of the existing works, or succinctness.
%
% \[
% Condensing\ \ =\ \ Compression\ \ \cup\ \ Succinctness\ \ \cup\ \ Relabeling\ \ \cup\ \ Summarization\ \ \cup\ \ Others
% \]

\macb{What Is the Scope of Existing Surveys?}
Existing surveys on graph compression cover a relatively small part of the domain.
Zhou provides a brief survey~\cite{zhou2015graph} with 14 references.
Maneth and Peternek~\cite{maneth2015survey}
%
%describe the total of 47 references and they 
%
cover a part of succint
representations, RDF graph compression, and several works categorized under
the common name ``Structural Approaches''.
Finally, there are other surveys that describe fields only partially related to
graph compression: compressing polygonal meshes~\cite{maglo20153d}, summarizing
graphs~\cite{liu2016graph}, and compressing
biological data~\cite{hosseini2016survey}.  We discuss these surveys in more
detail in the related parts of this work.

% one surveys graph summarization: a task related to graph compression in several
% ways but with the core difference related to the purpose, which is not reducing
% the size of a graph but to provide a description that best summarizes some
% selected properties of a graph. We will refer to all these works in more detail
% later. 

\section{Background}

We first present concepts used in all the sections and summarize the key 
symbols in Table~\ref{tab:symbols}.

\begin{table}[h!]
\vspace{-1em}
\centering
\footnotesize
%\scriptsize
%\ssmall
\sf
\begin{tabular}{ll@{}}
\toprule
                    $G, \mathbf{A}$ & A graph $G=(V,E)$ and its adjacency matrix; $V$ and $E$ are sets of vertices and edges.\\
                    $n,m$&Numbers of vertices and edges in $G$; $|V| = n, |E| = m$.\\
                    $d, \hat{d}, D$&Average degree, maximum degree, and the diameter of $G$, respectively.\\
                    $d_v, N_v$ & The degree and the sequence of neighbors of a vertex $v$.\\
                    $\mathscr{O}, \mathscr{A}, \mathscr{A}_v$ & \makecell[l]{Data structures for, respectively, the pointers to the adjacency data of each vertex,\\the adjacency data of a given graph $G$, and the adjacency data of vertex $v$.} \\
                    $N_{in,v}, N_{out,v}$ & The in-neighbors and out-neighbors of a vertex $v$.\\
                    $N_{i,v}$ & The $i$th neighbor of $v$ (in the order of increasing labels).\\
%                    \midrule
% \multirow{3}{*}{\begin{turn}{90}\shortstack{Various}\end{turn}}
%                      $T,W$ &The number of threads; work complexity of a given scheme.\\
%                      $C,n_c$ &Chunk height; the number of chunks in SlimSell or Sell-$C$-$\sigma$.\\
%                      $\sigma$ &Sorting scope in SlimSell and Sell-$C$-$\sigma$ ($\sigma \in [1,n]$).\\
\bottomrule
\end{tabular}
\caption{The most important symbols used in the paper.}
\vspace{-2em}
\label{tab:symbols}
\vspace{-0.5em}
\end{table}

\vspace{-0.5em}
\subsection{Graphs}

We model an undirected graph $G$ as a tuple $(V,E)$ (denoted also as $G(V,E)$);
$V$ is a set of vertices and $E \subseteq V \times V$ is a set of edges;
$|V|=n$ and $|E|=m$. If $G$ is directed, we use the name \emph{arc} to refer to
an edge with a specified direction. 
We consider both labeled and unlabeled graphs. If a graph is labeled, $V = \{1,
..., n\}$, unless stated otherwise.
%
% $A \subseteq V \times V $ denotes the set of all arcs. 
% 
$N_v$ and $d_v$ are the neighbors and the degree of a vertex $v$. The $i$th
neighbor of $v$ (in the order of increasing labels) is denoted as $N_{i,v}$;
$N_{0,v} \equiv v$. We use the name ``label'' or ``ID'' interchangeably.
% 
% The (non-negative) weight of an edge $(v,w)$ is
% {\small$\mathcal{W}_{(v,w)}$}. 
% 
We denote the maximum degrees for a given $G$ as $\hat{d}$, $\hat{d}_{in}$
(in-degree), and $\hat{d}_{out}$ (out-degree).
%
% The average degree is denoted with a bar ($\overline{d}$). 
%
$G$'s diameter is $D$.
%
% Each vertex $v$ is assigned a label $\lambda(v) \in \mathbb{N}$. Labels are
% unique ($v \neq u \Rightarrow \lambda(v) \neq \lambda(u)$), not necessarily
% contiguous, and can be freely reassigned. We identify vertices with their
% labels ($v \equiv \lambda(v)$) when it does not cause confusion.

% We focus on three fundamental queries~\cite{succinct_bound} on the graph
% structure: \textsf{deg(v)} and \textsf{neigh(v)} return the degree and the
% neighbors of a given vertex $v$; \textsf{adj(v,w)} returns \textsf{true} if two
% vertices $v$ and $w$ are neighbors, and \textsf{false} otherwise.

We also consider more complex graph properties: arboricity and genus.
Arboricity is a minimum
number of spanning forests that cover all the edges of a given graph. 
Next, a graph $G$ has a genus $g \ge 0$ if it can
be drawn without crossing itself (i.e., its edges) on the surface
of a sphere that has $g$ handles (\emph{such as in a coffee mug}).
For example, a graph with $g = 1$ can be drawn on a torus~\cite{thomassen1989graph}.

\subsection{Graph Representations}

$G$ can be represented as an \emph{adjacency matrix} (AM) or \emph{adjacency lists}
(AL). An AL consists of a contiguous array with the adjacency
data (denoted as $\mathscr{A}$) and a structure with offsets to the neighbors
of each vertex (denoted as $\mathscr{O}$). AL uses $\mathcal{O}(n \log m +
m \log n)$ bits while AM uses $\mathcal{O}\left(n^2\right)$.
\textsf{AL} needs $O\left(\hat{d}\right)$ time to check if two vertices
are connected while obtaining $N_v$ or $d_v$ takes $O(1)$ time. 
For \textsf{AM}, it takes $O(1)$ to verify if two vertices are connected and
$O\left(\hat{d}\right)$ to obtain $N_v$ and $d_v$.

% \textsf{AL} needs $O(\hat{d})$ time to answer \textsf{adj(v,$\cdot$)} while
% \textsf{neigh(v)} and \textsf{deg(v)} take $O(1)$ time. Fetching the offset of
% the adjacency data of vertex $v$ takes $O(1)$ time and is referred to as
% \textsf{off($v$)}.
% 
% \textsf{AM} is not commonly used because it uses $O(n^2)$ bits of space and, in
% its uncompressed form, offers storage advantages over \textsf{AL} only for very
% dense graphs where $2m \log n + n \log m > n^2$.  For \textsf{AM}, it takes
% $O(1)$, $O(\hat{d})$, $O(\hat{d})$, and $O(1)$ time to answer the \textsf{adj},
% \textsf{neigh}, \textsf{deg}, and \textsf{off} queries, respectively.

\subsection{Graph Families}
\label{sec:graph_families}

We next describe families of graphs used in 
the following sections; we provide them to make our work self-contained.
For simplicity, we focus on intuitive descriptions instead of formal definitions.
%
% \begin{itemize}[noitemsep, leftmargin=1.0em]
%
\macb{General graphs} can have multiple edges between the same two vertices (i.e., multigraphs) and loops (a loop
is an edge starting and ending at the same vertex).
\macb{Simple graphs} have no loops and no parallel edges.
\macb{Loop-free graphs} are general graphs without loops.
\macb{Stick-free graphs} do not have vertices of degree one.
\noindent
A \macb{transposed graph} has reversed arc directions of
the input directed graph~\cite{boldi2009permuting_social}.
\noindent
A \macb{graph embedding} into a certain surface is a drawing of a graph $G$ on
that surface so that $G$'s edges may intersect only at their
endpoints~\cite{bernhart1979book}.
\noindent
A \macb{planar graph} 
can be drawn on the plane so that vertices are the only place where the edges intersect. 
\noindent
An \macb{outerplanar graph} has a planar drawing such that all
vertices belong to the drawing of the outer face of the graph (intuitively,
all the vertices lie on the boundary between the graph and the outer part of the ``plane''
surrounding the graph).
\noindent
A \macb{plane graph} (also \macb{planar drawing} or \macb{plane drawing})
is a \macb{planar embedding} of a planar graph.
\noindent
A \macb{$k$-page graph $G$} has its edges partitioned into $k$ sets $E_1, ...,
E_k$ so that, for each $i$, a graph $G(V,E_i)$ is planar~\cite{kannan1985unraveling}.
% 
% $G(V,E_i)$ is a graph with a vertex set $V$ and edge set $E_i$
%
In a drawing of a graph $G$ on $k$ pages each vertex of $G$ constitutes a point
on a ``spine'' of the book formed by these $k$ pages and each edge of $G$ is
drawn as a curve within a single page.
\noindent A \macb{$k$-page graph embedding} on a given surface is a $k$-page
drawing of a graph $G$ on this surface without any edge crossings.
\noindent A \macb{$k$-connected graph $G$} (also called $k$-vertex connected) is
such a graph in which one cannot find a set of $k-1$ vertices whose removal
disconnects $G$.
%
% \end{itemize}
%
\noindent A \macb{map}
is a \emph{``topological equivalence class of planar embeddings of planar
graphs''}~\cite{keeler1995short}. 
%
%In
%encoding a map we are required to encode the topology of the embedding, i.e.,
%incidences among faces, edges, and vertexes, as well as the graph. 
%
Because \emph{``a map is an embedding of a unique graph, but a given graph may
have more than one embedding''}~\cite{keeler1995short}, a map usually requires
more bits to encode than a corresponding graph. 
A \macb{plane triangulation} is a plane graph where each face consists of
exactly three edges; a plane triangulation may contain self-loops and multiple
edges.
An \macb{Erdős–Rényi graph} $\mathcal{G}(n,p)$~\cite{erdos1961evolution} is a
graph with a uniform degree distribution where $n$ is the number of vertices
and $p$ is a probability that an arbitrary edge is present in the graph
(independently of other edges).
A \macb{separable graph} is a graph 
in which we can divide $V$ into two subsets of vertices of approximately
   identical size and with the size of a vertex cut
  between these two subsets being asymptotically smaller than $|V|$.
Other considered graph classes are \macb{graphs with bounded arboricity}
(intuitively, they are
\emph{uniformly sparse})
and
\macb{graphs with bounded genus}.

% \subsection{Graph Theory Notions}
% 
% An \emph{automorphism} of a $G$ is a permutation of the vertices of $G$ that
% preserves the adjacency structure. The collection $\text{Aut}(G)$ of all $G$'s
% automorphisms is referred to as the automorphism group of $G$.

\begin{figure*}%[!h]
\centering
\includegraphics[width=1.03\textwidth]{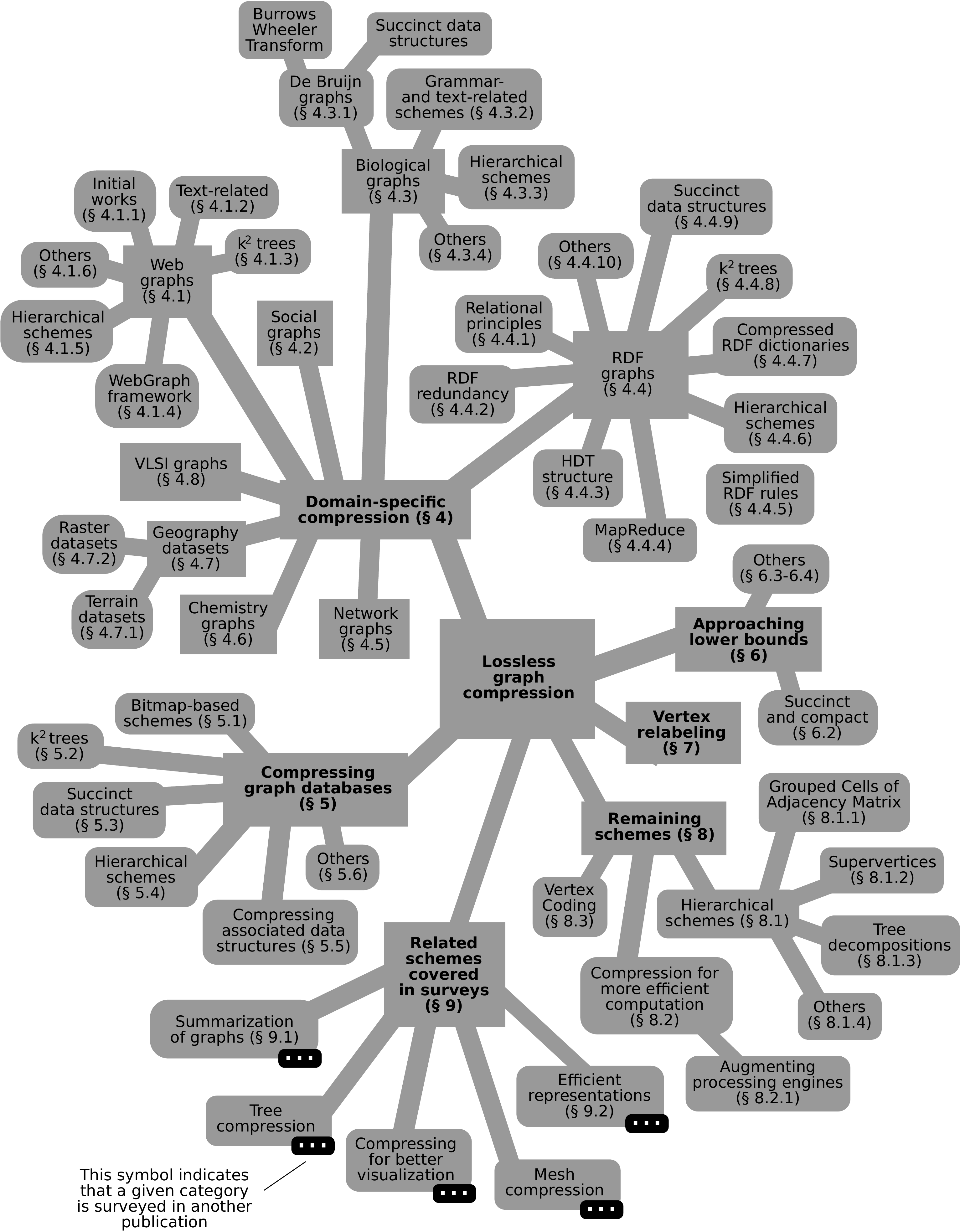}
\caption{(\cref{sec:categorization}) The categorization of the considered domains of lossless graph compression.}
\label{fig:categorization}
\end{figure*}

\subsection{Codes}

Finally, we summarize codes for encoding integers that are used in various
works to encode adjacency arrays of vertices.
Elias $\gamma$~\cite{elias1975universal} is a universal code for positive
integers. It is often used when the maximum possible number to be encoded
cannot be determined beforehand; when using this code, the size of a value $x$
is $2 \lceil \log x \rceil +1$ [bits]. 
Elias $\delta$~\cite{elias1975universal} is a universal and asymptotically
optimal code for positive integers. With this code, the size of a number $x$ is
$\lceil \log x \rceil + 2 \lceil  \log \lceil \log x \rceil + 1 \rceil + 1$
[bits].
Elias-Fano~\cite{vigna2013quasi, ottaviano2014partitioned} is a recent development for monotonically 
increasing sentences of integers.
%
% Elias Omega~\cite{elias1975universal} & - & Universal code for positive integers.\\
%
% \item \macb{Elias Fano~\cite{x}}
%
Golomb~\cite{golomb1966run} is an optimal non-universal prefix code for
alphabets following a geometric distribution. The code is suitable for
sequences of integers where small values are much more likely to occur than
large values.
% %
% \item \macb{Nybble~\cite{x}}
% %
% \item \macb{Rice~\cite{x}}
% %
Gray code~\cite{gray1953pulse} is an arrangement (i.e., a total ordering) of
numbers (or other entities such as vectors) such that the binary
representations of consecutive numbers or any other entities differ by exactly
one bit.
% %
% \item \macb{Varint~\cite{dean2009challenges}}
%
$\zeta$~\cite{boldi2005codes} is a code suited for integers that follow the
power law distribution with the exponent smaller than two.
Finally, $\pi$~\cite{apostolico2009graph} is a universal code suited for
integers that follow the power law distribution with the exponent close to one.

\section{Taxonomy and Domain Dimensions}

In this section, we describe how we categorize existing work in this survey.
Figure~\ref{fig:categorization} depicts the
hierarchy of the considered domains. 

% and the space of the considered schemes.

\vspace{-0.5em}
\subsection{How Do We Categorize Existing Work?}
\label{sec:categorization}

Graph compression is related to various areas such as databases or information theory.
Schemes in these areas often share various common
features, for example addressing static or dynamic graphs. This poses a
question on how to categorize the rich world of graph compression studies to
enable its systematic analysis.
%
% We provide an overview of our categorization in Figure~\ref{fig:space}.
%
In the following, we dedicate a separate section (\cref{sec:specific-domains}--\cref{sec:remaining}) to one particular \emph{area of research} such as
compressing graph databases. Thus, we devote one section to the work done
within one specific community. Such community-driven areas constitute \emph{the
first dimension} of our categorization.  We describe how we identify these
areas in~\cref{sec:areas}.
Second, different areas of graph compression may use the same
\emph{techniques} for compressing graphs, or various techniques may be used in
one publication or algorithm. For example, it is common to combine \emph{gap
encoding} and \emph{variable-length codes} to compress web graphs. \emph{The
second dimension} in our categorization are thus \emph{techniques} that reduce
the size of graphs. We dedicate~\cref{sec:existing-techniques} to describe the
most important techniques. 
Finally, applications of a given technique within a certain area may have
different \emph{features}. For example, one can use a specific technique for
either static or dynamic graphs. Consequently, the third dimension are
features; we describe them in~\cref{sec:features}.
%
% However, to enable a systematic overview of differences and similarities
% between the presented schemes, we hereby illustrate various aspects of graph
% compression and we propose to treat these aspects as dimensions of a
% multidimensional space where each compression scheme can be seen as a point
% in this space. Such point is defined by particular values of the given set of
% features (dimensions). An example dimension is \emph{compression type} with
% two possible values: lossless and lossy.

We also discuss the existing categorizations and taxonomies
in~\cref{sec:existing-taxonomies}.

% \begin{figure}[!h]
% \centering
% \includegraphics[width=0.75\textwidth]{pictures/space.eps}
% \caption{(\cref{sec:categorization}) The space of the considered lossless graph compression schemes.}
% \label{fig:space}
% \end{figure}

\subsubsection{Areas}
\label{sec:areas}

Many papers are dedicated to {compressing graphs from specific domains} such as
web graphs, biological networks, social graphs, and others; we describe them
in~\cref{sec:specific-domains}. Second, we describe works related to
compressing graph databases (\cref{sec:databases}).  Next, various schemes that
we list in~\cref{sec:bounds} are devoted to approaching the storage lower
bounds while ensuring fast (ideally constant time) queries.  Moreover, we
discuss optimization approaches to improve \emph{graph layouts}, which
ultimately reduces space occupied by a graph.
Finally, we devote a separate section for various works that cannot be
categorized in one of the above (see~\cref{sec:remaining}) and to areas related
to graph compression and covered in other surveys
(see~\cref{sec:others_surveyed}).

% \[
% Condensing\ \ =\ \ Compression\ \ \cup\ \ Succinctness\ \ \cup\ \ Relabeling\ \ \cup\ \ Summarization\ \ \cup\ \ Others
% \]

\subsubsection{Techniques}
\label{sec:existing-techniques}

We now briefly present several common techniques used in various areas, as well as
examples of their usage, to
improve the clarity of the survey. 
% 
% We later (\cref{sec:techniques_taxonomy})
% present a detailed taxonomy of techniques discussed in this survey.

\vspace{1em}
%\begin{itemize}[noitemsep, leftmargin=1.0em]
%
%\item 
\noindent
\macb{Variable-Length Encoding} In this technique, vertex IDs stored in
the adjacency array are encoded with one of the selected variable-length codes
such as Varint.

\begin{figure}[!h]
\vspace{-0.5em}
\centering
\includegraphics[width=1.0\textwidth]{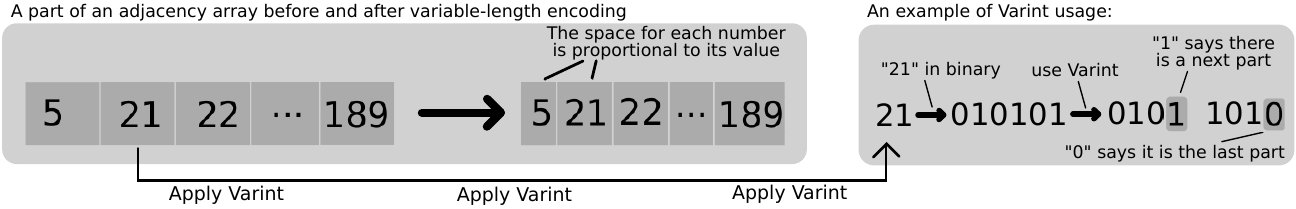}
\vspace{-2.5em}
\caption{An example of variable-length encoding.}
\label{fig:var-len_enc}
\end{figure}

%\item 
%
\noindent \macb{Vertex Relabeling} The main idea is to change the initial IDs
of vertices so that the new IDs, when stored, use less space. We also use the
name \emph{vertex permutations} to refer to this technique. This scheme is
usually combined with variable-length encoding.

\begin{figure}[!h]
\vspace{-0.5em}
\centering
\includegraphics[width=0.6\textwidth]{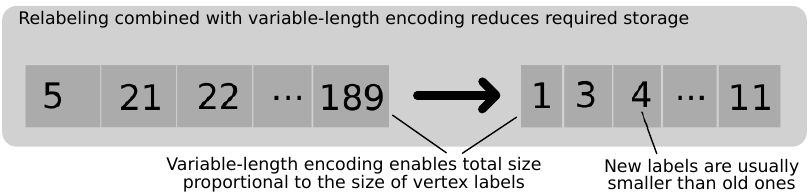}
\vspace{-1em}
\caption{An example of vertex relabeling combined with variable-length encoding.}
\label{fig:var-len_enc}
\end{figure}

%\item 
\noindent
\macb{Reference Encoding} Here, identical sequences of vertices in the
adjacency arrays of different vertices are identified. Then, all such sequences
(except for a selected one) are encoded with references~\cite{adler2001towards,
randall2002link}. One can implement reference encoding with \emph{copy lists}:
sequences of 1s and 0s that indicate whether or not a given number is retained
in the current adjacency array.

\begin{figure}[!h]
\vspace{-0.5em}
\centering
\includegraphics[width=0.5\textwidth]{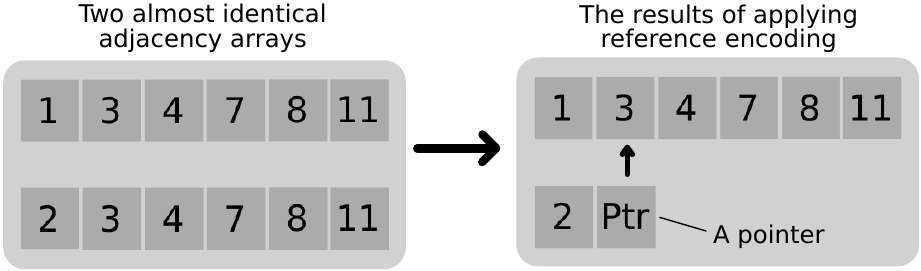}
\vspace{-1em}
\caption{An example of reference encoding (no copy lists).}
\vspace{-1em}
\label{fig:var-len_enc}
\end{figure}

%\item 
\macb{Run-length Encoding} This scheme enhances copy
lists in reference encoding. The key idea is to provide the size of
consecutive sequences of 1s and 0s instead of the actual 1 and 0
values~\cite{asano2003compact}.

%\item 
\macb{Huffman Degree Encoding} The core idea in this scheme is to use
fewer bits to encode vertex IDs of higher degrees. Thus, $|\mathscr{A}|$ is
reduced as vertex IDs that occur more often use fewer bits.

%\item 
\macb{Log Encoding} This scheme uses $\lceil \log n \rceil$ bits to
encode each vertex ID in a graph with $n$ vertices.  

%\item 
\macb{Interval Encoding} Here, consecutive vertex IDs (e.g., $x, x+1,
..., x+k$) are stored using the interval boundaries $x$ and $x+k$.

%\item 
\macb{Gap Encoding} This scheme preserves differences between
vertex IDs rather than the IDs themselves. The motivation is that, in most
cases, differences occupy less space than IDs. Several variants can be used
here; the most popular is storing differences between the IDs of the
consecutive neighbors of each vertex $v$, for example $N_{1}(v) - v, N_{2}(v) -
N_{1}(v), ..., N_{d_v-1}(v) - N_{d_v-2}(v), N_{d_v}(v) - N_{d_v-1}(v)$ (the
first of the above differences is sometimes called an \emph{initial distance}
and each following: an \emph{increment}). Assuming each $\mathscr{A}_v$ is
sorted, one must use an additional bit to indicate the sign of the first
difference. Another variant stores the differences between $v$ and each of its
neighbors: $N_{1}(v) - v, N_{2}(v) - v, ..., N_{d_v-1}(v) - v, N_{d_v}(v) -
v$.

%\end{itemize}

There are many more techniques used in various areas of graph compression.
We defer their description to the relevent parts of the survey. These are,
among others, $k^2$-trees and their variants, hierarchical schemes based
on the concept of supervertices and superedges, and schemes that reorder
the rows or columns of the graph adjacency matrix.
Some techniques are used in various areas but they are themselves actively
developed and they constitute a separate area of graph compression. Examples
are succinct data structures or schemes that relabel vertices.

% $k^2$-trees and related
% 
% hierarchical 
% - supervertices (also called supernodes) + superedgres
% -- clusters
% -- same neighborhoods
% - succinct 
% - compact
% 
% reordering matrix
% 
% relabveling

\subsubsection{Features}
\label{sec:features}

We briefly present several features used in various compression areas to
improve the clarity of the survey. We later (\cref{sec:features_taxonomy})
discuss selected features in more detail.

\begin{itemize}[noitemsep, leftmargin=1.0em]
%
% \item \macb{Compression Type} The type can be either lossless or lossy. Some
% schemes offer both~\cite{sigcom-stuff} types, for example by providing a
% parameter that specifies the maximum allowed error; setting this parameter to 0
% would ensure lossless compression.
%
\item \macb{Graph Dynamicity} This feature indicates whether a graph to be
compressed is assumed static or dynamic (and thus allowing for any changes to
its structure or labels).
\item \macb{Problem-Awareness} This feature determines whether a given compression
schemes is tuned to some specific algorithm or graph problem to be
solved over the compressed graph.
\item \macb{Graph-Awareness} Here, we determine whether a given scheme is tuned
(or designed) for some specific graph classes or whether it works for generic
graphs.
\item \macb{Streaming Graphs} This feature indicates whether a given scheme
addresses graphs that are processed as a stream of edges.
\end{itemize}

\subsection{Existing Categorizations}
\label{sec:existing-taxonomies}

We also survey the existing categorizations of graph compression schemes.
First, Boldi et al.~\cite{boldi2009permuting_social} indicate that permutations that relabel
vertices can be \emph{intrinsic} (also called \emph{coordinate-free}) or
\emph{extrinsic}. The former relabel IDs basing only on the
graph structure (i.e., vertices and edges). The latter also rely on some
additional information such as URLs. 
Next, Dhulipala et al.~\cite{dhulipala2016compressing} identify three
categories of schemes for graph (and index) compression. First, there are
\emph{structural} approaches that collapse frequent graph subgraphs (such
as cliques). Second, there exist schemes that encode the set of graph edges 
represented with a sequence of integers. Finally, various works propose vertex
relabelings (label permutations) that minimize a given metrics, for example the
sum of differences between consecutive neighbors in each adjacency list.
Another study~\cite{nourbakhsh2014matrix} distinguishes between
\emph{structural} approaches and the ones based on using the notion of
entropy~\cite{shannon1998mathematical}.
Finally, schemes for compressing web graphs use the notions of \emph{locality} and
\emph{similarity}~\cite{boldi2010compressing}. Locality indicates that most links
from page~$x$ lead to pages on the same host (that often share a long path
prefix with $x$).  Similarity means that pages from the same host have many
links in common.

% \cite{asano2008efficient}: locality types.

% \maciej{Sort these out:}
% %
% (1) locality is taken advantage of by not writing the successors
% of a node explicitly, but as gaps (i.e., as differences
% between each successor and the previous one), using
% suitable variable-length encodings that write few bits
% for short numbers;
% %
% (2)  similarity is exploited using differential compression.
% More precisely, every successor list (i.e., every matrix
% row) is compared with a number of previous rows,
% and the BV algorithm tries to compress it differentially
% with respect to (i.e., as a difference from) each
% such row, choosing at the end the row that gave the
% best compression (or none, if representing the row nondifferentially
% gives better compression); the number
% of rows considered in this process is called the window
% size (hereafter denoted by W). Of course, larger
% window sizes produce slower compression, but usually
% guarantee better ratios: the default is W = 7.

\section{Compressing Graphs in Specific Domains}
\label{sec:specific-domains}

A large portion of research in graph compression is dedicated to compressing
graphs in some specific domains. We now present the related efforts.

% \macb{Features}
% %
% Unless stated otherwise, the described schemes operate on static graphs.

% \subsection{Background Concepts and Notions}
% 
% \subsubsection{Alphabet and Text}
% 
% A \emph{text} is a sequence of $n$ symbols $T = t_1 ... t_n$ over an alphabet
% $\Sigma$ that contains all the symbols that can be used to construct $T$.

\subsection{Web Graphs}
\label{sec:web_graphs}

We start with web graphs. 

\subsubsection{Initial Works}

The first works on web graph compression often use various combinations of techniques
such as Huffman degree encoding, log encoding, gap encoding, differential encoding,
and various variable-length codes.
This domain was opened by the work on the Connectivity
Server~\cite{bharat1998connectivity}, a system for storing the linkage
information found by the AltaVista~\cite{silverstein1999analysis} search
engine. Connectivity Server associates each URL with an integer, sorts these
integers (in each adjacency data structure $\mathscr{A}_v$) according to the
URL lexicographic order, and uses gap encoding on the integers.
 
Another early work analyzes the web graph structure~\cite{broder2000graph} and
also enhances the original compression scheme in the Connectivity Server.
Then, Wickremesinghe et al.~\cite{wickremesinghe2000link} uses Huffman codes to
encode references to links in the Connectivity Server.
Moreover, Adler and Mitzenmacher~\cite{adler2001towards} propose to compress
web graphs with a Huffman-based scheme applied to in-degrees. They also use
reference encoding and log encoding.

Suel and Yuan~\cite{suel2001compressing} compress URLs using common prefixes.
To compress links, they distinguish between (1) global links 
connecting different hosts and (2) local links connecting sites on the same host. 
First, they find a selected number of highest in-degree URLs and encode
\emph{global} links to the URLs with a Huffman code. For other global
links, they use log encoding or encode the link
with a Golomb code, depending on the out-degree of a given URL. Next,
they encode local links between highest degree sites within the same host using a Huffman code.
%

% they identify two local link classes. For each host~$h$, they determine the
% given number of highest degree sites for local links within $h$. Links to
% these pages are encoded with a Huffman code. 

The Link Database~\cite{randall2002link} is a system for storing web graphs. It
uses various techniques (delta codes, Huffman codes, Gray orderings, and nybble
codes); the related analysis identifies two important web graph properties,
namely \emph{locality} and \emph{similarity}. The Link Database compresses
offset arrays $\mathscr{O}$ by using different bit counts to store offsets for
different degree ranges. For example, it uses a 32-bit index to the start of a
given range of vertices, and then only 8-bit offsets for each of the following
vertices. 
% 
% Lexicographic codes for URLs are also used. Finally, reference encoding is
% also incorporated. 

% In Link2, each adjacency list was sorted in ascending order and represented
% by the list of the delta values of its elements. The delta value of the first
% element of the list is its difference from the source of the link, and the
% delta value of each of the other elements is the difference between its
% previous element and itself. They observed that the delta values tended to be
% close to zero, resulting from a kind of locality of the web graph that the
% destinations of the links originating at the same page are often near from
% each other. According to this observation, they used variable-length nybble
% codes to represent the delta values in their encoding of the web graph. By
% combining this representation with other techniques, Link2 encoded a graph in
% 11.03 bits per edge on average.

\subsubsection{Text-Related Works}

Next, we present efforts that revolve around treating the input graph $G$ as text and
using the associated compression methods.
Navarro~\cite{navarro2007compressing} proposes to regard $G$ as text
and to utilize existing techniques for text compression and indexing. Specifically,
he uses the \emph{Compressed Suffix Array (CSA)}~\cite{sadakane2003new}
structure as a basis for his graph representation. CSA is a \emph{compressed
full-text self-index}: a data structure constructed over some text $T = t_1 ... t_n$
(over an alphabet $\Sigma$). CSA has the size proportional to the size of the
text being compressed and simultaneously enables accessing any substring of $T$ as well as 
full $T$ itself. Therefore, storing $T$ becomes unnecessary and still some search
operations on it are enabled. Now, the key idea is to treat an input graph $G$ as
$T$ and use various CSA's functionalities to enable efficient accesses of $G$ 
without the need for decompression.

\begin{figure}[!h]
\vspace{-0.5em}
\centering
\includegraphics[width=1.0\textwidth]{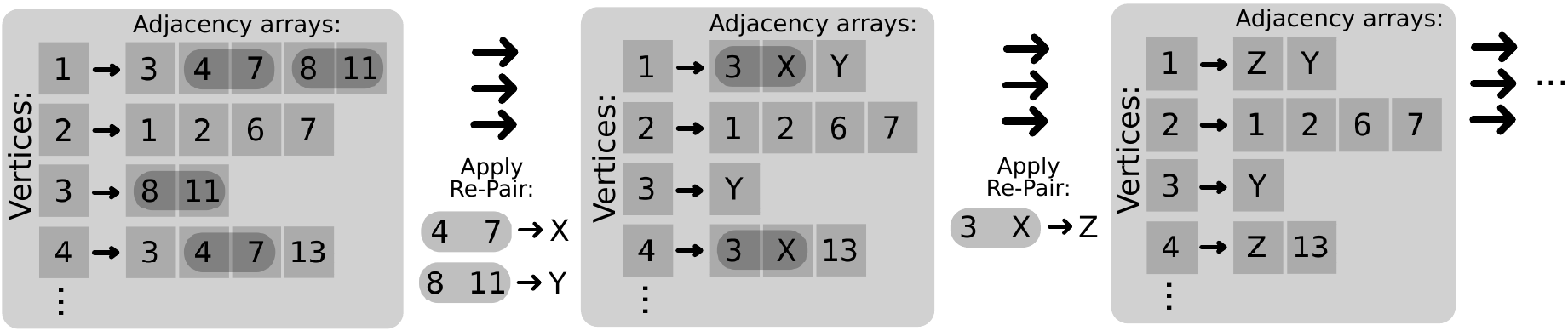}
\vspace{-2em}
\caption{Several example iterations of the Re-Pair scheme.}
\vspace{-0em}
\label{fig:repair}
\end{figure}

Claude and Navarro~\cite{claude2007fast} use
Re-Pair~\cite{larsson2000off}: a phrase-based compression scheme that enables fast decompression that is also 
\emph{local}: does nor require accessing the whole graph. Re-Pair
repeatedly finds the most frequent pairs of symbols in a given graph representation 
and replaces them with new symbols. This is repeated as long as storage is reduced.
An example is in Figure~\ref{fig:repair}.
% 
% They first design an approximate Re-Pair scheme that: (1) works on an arbitrary
% sequence, (2) uses as little memory as desired on top of the input, (3) can
% trade accuracy for speed, and (4) is able to use secondary memory.  The main
% idea of Claude and Navarro is to apply this variant of Re-Pair over
% $G$'s adjacency array~$\mathscr{A}$.

The approach based on Re-Pair was further extended by Claude and
Navarro~\cite{claude2010extended}. They combine it with an approach based on
perceiving $G$ as a binary relation (on $V \times V$). Next, they use several schemes 
developed specifically for answering queries over binary
relations~\cite{barbay2007adaptive, barbay2007succinct}.  The motivation for
such a combination is to derive \emph{both} $N_{out,v}$ \emph{and} $N_{in, v}$
fast.

\subsubsection{$k^2$ Trees}
\label{sec:k2_web}

\begin{figure*}%[!h]
\vspace{-0.5em}
\centering
\includegraphics[width=1.0\textwidth]{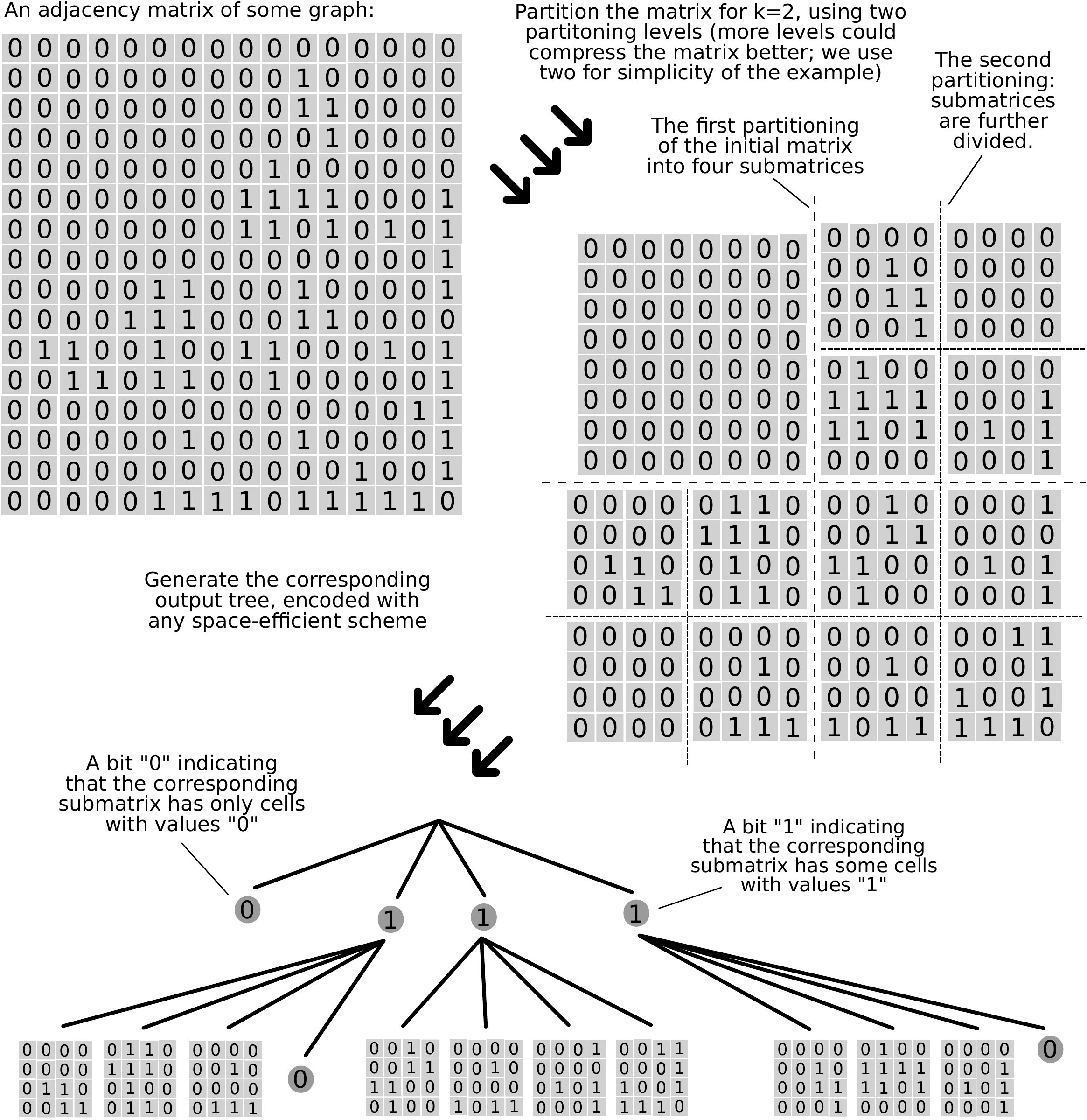}
\vspace{-2em}
\caption{An example application of the $k^2$ tree scheme.}
\vspace{-1em}
\label{fig:k2}
\end{figure*}

We now describe efforts related to so called $k^2$ trees.
In their seminal work, Brisaboa et al.~\cite{brisaboa2009k2,
gonzalez2011algorithms} present a graph representation 
where a graph is modeled with a tree. The structure of the tree
reflects the structure of the graph adjacency matrix $A$.
% 
% that uses a tree structure that takes advantage of the structure of the
% adjacency matrix $A$ of web graphs. 
% 
%
An example is presented in Figure~\ref{fig:k2}.
Initially, $A$ is divided into $k^2$ submatrices of identical size ($k$ is a
parameter); these submatrices are recursively divided in the same way. Now, the
key idea is to represent $A$ as a $k^2$-ary tree (called a $k^2$ tree) that
corresponds to the above recursive ``partitioning'' of $A$.  Each internal tree
node has $k^2$ children.  Each tree node stores one bit of data. At every
partitioning level, if a given submatrix to be partitioned contains only 0s,
the corresponding tree node contains 0. Otherwise, it contains a 1. The
resulting tree is encoded using a special simplified tree encoding that ensures
asymptotically low compression ratio~\cite{rahman2006engineering}. 
The worst case size of a $k^2$ tree is $k^2 m (\log_{k^2} \frac{n^2}{m} + \mathcal{O}(1))$ bits.
The $k^2$
representation ensures obtaining $N_{out,v}$ and $N_{in,v}$ fast ($\mathcal{O}(\sqrt{m})$ worst
case time, assuming $m$ ones are uniformly distributed in the adjacency matrix).
%
% 
% it also
% supports accessing other graph ele not usually considered in compressed graph
% representations. 
%
Finally, high compression rations obtained with $k^2$ trees are due to the fact that
these trees directly utilize the sparseness and clustering properties of $A$.

Claude and Ladra~\cite{claude2011practical} further extended the $k^2$ tree
idea by combining it with the Re-Pair scheme. Specifically, they first split
the graph into subgraphs that correspond to different web domains. After that,
the key idea is to encode each subgraph with a $k^2$ tree representation and
encode the remaining inter-subgraph edges with Re-Pair. 
Furthermore, a study on the best data distribution for query processing on
graphs compressed with $k^2$ trees can be found in the work by Alvarez et
al.~\cite{alvarez2013distributed}.
The work on $k^2$ trees was also extended by Brisaboa et
al.~\cite{brisaboa2014compact} to ensure obtaining some properties of the graph
structure fast. For example, they accelerate finding the successors of a
certain web page that are within a given range of web pages. Moreover,
additional enhancements ensure better compression ratios.  First, they vary $k$
across the tree levels: larger $k$  is used in higher tree levels while smaller
$k$ is used in last tree levels. This enhancement uses the observation that
lower tree levels require fewer bits to cover 1s (i.e., the adjacency matrix in
the considered graphs is usually very sparse and the existing 1s can be covered
by square submatrices of small dimensions, thus the $k$ parameter can be
assigned low values for the last levels of respective $k^2$ trees). Second,
they use multiple $k^2$ trees for one adjacency matrix by partitioning the
matrix and building one tree per partition independently of others. This
decreases construction times and enables better adjustment between the
granularity of the tree ($k$) and the submatrix sparsity. Third, the last level
is additionally compressed with a variant of Huffman encoding: respective
submatrices are sorted by frequency and are assigned numerical values that
identify (and point to) each submatrix; lower numbers are assigned to more
frequent submatrices to lower the total number of bits used to represent the
matrices.

\subsubsection{WebGraph Framework}
\label{sec:web_graphs_web_graph}

We devote a separate subsection to describe the efforts related to the WebGraph
Framework, a mature framework targeted at compressing web graphs.

Boldi and Vigna provide several works on compressing Web graphs. First, they
developed the WebGraph framework~\cite{boldi2004webgraph} that is a freely
available suite of codes, algorithms, and tools for compressing graphs with a
special focus on web graphs.
In addition to exploiting the two well-known web graph properties (locality and
similarity) they also identify three further properties: (1) ``similarity
concentration'', i.e., either two adjacency lists share almost nothing,
or they are characterized by large overlap, (2) ``consecutivity is
common'', i.e., many links within a page are consecutive with respect to the
lexicographic order, and (3) ``consecutivity is the dual of distance-one
similarity'', i.e., if there is a lot of similarity between vertices and their
respective successor lists in lexicographical ordering, the transpose of the
given graph must contain large intervals of consecutive links.
Now, WebGraph uses the following compression techniques: (1) gap encoding, (2)
reference encoding, (3) differential encoding, and (4) interval encoding.
Reference encoding may have an arbitrary number of levels, bounded by a user
parameter. Another parameter controls the minimum number of integers that
enables a given sequence of numbers to be taken into consideration when
applying reference encoding. Finally, for high performance, adjacency lists are
built \emph{lazily} (i.e., the data available through reference encoding is
only fetched when required). 

Moreover, Boldi et al.~\cite{boldi2009permuting_social}
test several existing vertex permutations. They also propose two new
permutations based on the Gray~\cite{knuth2005art} ordering. The key idea is the 
permutation of the adjacency matrix rows so that the adjacent rows change
according to the Gray code. Moreover, they analyze transposed graphs and
illustrate that coordinate-free permutations improve compression rates
(\emph{coordinate-free} permutations deliver approximately the same
compression ratios regardless of how the vertex initial ordering). They conclude
that the Gray ordering may improve the compression rates and that
coordinate-free orderings are particularly efficient for transposed graphs (in
some cases reaching the level of 1 bit per edge). 

Boldi et al.~\cite{boldi2011layered} conduct an analysis that aims to formally
understand why the existing approaches compress web graphs well.  For this,
they first observe that it is important for high compression ratios to use an
ordering of vertex IDs such that the vertices from the same host are close to
one another. To understand this notion more formally, they propose
\emph{measures} of how well a given vertex ordering $\pi$ respects the
partitioning $\mathscr{H}$ due to hosts.
Specifically, they first model the host transition (HT) probability: a fraction
of vertices being followed (when considering the $\pi$ ordering) by a vertex
hosted elsewhere:

\vspace{-1em}
\small
\begin{gather}
HT(\mathscr{H},\pi) = \frac{1}{n} \cdot \sum_{i=1}^{n-1} \delta \left( \mathscr{H}\left[\pi^{-1}(i)\right], \mathscr{H}\left[ \pi^{-1}(i-1)\right] \right)
\end{gather}
\normalsize

\noindent
Here, $\delta$ is the regular Kronecker's delta. $\mathscr{H}[x]$ denotes
the equivalence class of a vertex~$x$. This class is the set of vertices with the same host as $x$.

Moreover, the second used measure is adapted from work by Meil\v{a}~\cite{meila2005comparing}
and is called the Variation of Information (VI); it enables comparing two partitions
$\mathscr{H}$ and $\mathscr{H}_\pi$ that are associated with the original vertex
ordering and the $\pi$ ordering.
First, VI uses a notion of entropy $H(\mathscr{P})$ associated with a host partitioning
$\mathscr{P}$:

\vspace{-1em}
\small
\begin{gather}
H(\mathscr{P}) = -\sum_{S \in \mathscr{P}} P(S) \log (P(S)) 
\end{gather}
\normalsize
where $P(S) = \frac{|S|}{n}$.
Then, the \emph{``mutual information between two partitions''}~\cite{boldi2011layered} is defined as

\vspace{-1em}
\small
\begin{gather}
I(\mathscr{P}, \mathscr{T}) = \sum_{S \in \mathscr{P}} \sum_{T \in \mathscr{T}} P(S,T) \log \frac{P(S,T)}{P(S) P(T)}
\end{gather}
\normalsize
where $P(S,T) = \frac{|S \cap T|}{n}$. Finally, they define the VI measure as

\vspace{-1em}
\small
\begin{gather}
VI(\mathscr{P},\mathscr{T}) = H(\mathscr{P}) + H(\mathscr{T}) - 2I(\mathscr{P},\mathscr{T})
\end{gather}
\normalsize
\vspace{-1em}

Now, substituting $\mathscr{P}$ and $\mathscr{T}$ with $\mathscr{H}$
and $\mathscr{H}_\pi$ and observing that $I(\mathscr{H},\mathscr{H}_\pi) = H(\mathscr{H})$
gives

\vspace{-1em}
\small
\begin{gather}
VI(\mathscr{H},\mathscr{H}_\pi) = H(\mathscr{H}_\pi) - H(\mathscr{H})
\end{gather}
\normalsize
\vspace{-1em}

the authors use the HT and VI measures to compare various vertex orderings used
for web graphs.  In the second part of the paper, they show that their ordering
  called Layered Label Propagation (explained in more detail
  in~\cref{sec:social_networks} dedicated to social networks as it was
  introduced mostly in the context of social networks) outperforms other
  orderings proposed in the literature.

Other works due to Boldi, Vigna, and others include developing complete
instantaneous $\zeta$ codes for integers distributed as a power law with the
exponent smaller than two~\cite{boldi2005codes}, and presenting how to implement
WebGraph with Java~\cite{boldi2004webgraph_java}. Some benchmarks
about Web Graph can also be found in several empirical
analyses~\cite{giatsoglou2010managing, lashtaghani2014large}.

\subsubsection{Hierarchical Schemes}
\label{sec:hierarchical_web}

We now survey methods where a given part of the input graph (e.g., a clique
or a subgraph) is collapsed into a smaller entity (e.g., a vertex) to
ultimately reduce space. 

Raghavan and Garcia-Molina~\cite{raghavan2003representing} propose a
representation called S-Node.  They find and collapse subgraphs into
supervertices and then use Huffman encoding on in-degrees as well as reference
encoding. Edges between supervertices are merged and form superedges, with
information on which single edges should be added or removed to restore the
original graph structure.  To further enhance the representation, they
incorporate various types of graph partitioning and they utilize contiguous
vertex IDs with respect to their orderings in supervertices. 

Buehrer and Chellapilla~\cite{buehrer2008scalable} generate \emph{virtual
nodes} from frequent itemsets in the adjacency data (i.e., dense subgraphs are
replaced with sparse ones). For example, consider a fully connected directed
bipartite subgraph where $n_1$ and $n_2$ denote the numbers of vertices in each
of the two color classes; all the edges are directed towards one color class. The
proposed scheme collapses the edges in such a subgraph by introducing a virtual
node $v$. Now, $n_1$ edges from one class point to $v$ and $n_2$ edges point
from $v$ to each vertex in the other color class.  This reduces the edge count
from $n_1 n_2$ to $n_1 + n_2$ and is able to achieve a high compression ratio as
bipartite subcliques are frequent in web graphs~\cite{kumar1999trawling}.  In
this work, the complexity of the mining phase (i.e., where itemsets are found)
is bounded to $O(m \log m)$. To ensure this complexity
they first cluster similar vertices in the graph (vertices are
similar if a significant portion of their outlinks point to the same
neighbors). Next, they find patterns in the clusters, remove the patterns, and
replace these patterns with virtual nodes. 
Buehrer and Chellapilla's work was extended by Mondal~\cite{mondal2015real}
by providing insights into which values of respective algorithm
parameters are best suited in a given scenario.

Karande et al.~\cite{karande2009speeding} complement the work of Buehrer and
Chellapilla~\cite{buehrer2008scalable}. They show that it is possible to run various web graph
algorithms on graphs compressed with a scheme using virtual nodes so
that \emph{``their running times depend on the size of the compressed graph rather than 
the original}''~\cite{karande2009speeding}. They consider various algorithms, including schemes for link analysis, for assessing the
cardinalities of vertex neighborhoods, and various schemes based on random walks or matrix–vector
multiplication (PageRank~\cite{brin2012reprint},
HITS~\cite{kleinberg1999authoritative}, and SALSA~\cite{lempel2000stochastic}).
For example, they show how to transform PageRank results for a compressed
graph in order to derive the corresponding results for the original graph dataset.

Khalili et al.~\cite{khalili2009web} relabel vertices so that similar vertices
have closer IDs.  Second, they group similar vertices together and collapse
edges between groups into single \emph{superedges}. To keep track of the
collapsed edges they use an auxiliary data structure.

Anh and Moffat propose a hierarchical scheme for compressing web
graphs~\cite{anh2010local}. The key idea is to partition the adjacency arrays
into groups of $h$ consecutive arrays. Then, sequences of consecutive integers
in each of the $h$ arrays are replaced with new symbols to reduce
$|\mathscr{A}|$; this can be seen as grammar-based compression conducted
in a local manner~\cite{grabowski2011merging}. 

Grabowski and Bieniecki~\cite{grabowski2010tight, grabowski2011merging} present
two schemes; the first one offers higher compression
ratios while the second one is faster. The first scheme extends the reference
encoding from the WebGraph framework. Among others, it extends the binary
format of reference encoding by using more than two values to indicate more
possible referenced values. Another one creates blocks of $h$ adjacency lists.
Then, within each such block, $h$ adjacency lists are merged and then all
duplicate numbers in each merged sequence of lists are removed. Simultaneously,
every number receives an associated list that indicates which neighborhoods it
originally belongs to.

Hernandez and Navarro~\cite{hernandez2011compression_web_social} combine
several techniques to accelerate obtaining various graph properties and to
further reduce the required storage. First, they combine reducing the number of
graph edges (through virtual nodes~\cite{buehrer2008scalable}) with vertex
reordering~\cite{apostolico2009graph, boldi2009permuting_social, boldi2004webgraph} to
take advantage of locality and similarity combined with ordering and
interval/integer encoding.  Next, they combine $k^2$
trees~\cite{brisaboa2009k2} with virtual nodes as
well as several other schemes, e.g., Re-Pair~\cite{claude2007fast}.
They address web and social graphs. They first analyze various existing
compression methods for web graphs and show that a combination of schemes that
collapse edges and reorder vertices can offer outstanding compression ratios
and performance of graph accesses.  Second, they propose a novel compression
method that uses compact (as defined in~\cref{sec:bounds}) data structures to
represent social communities.

In other studies, Hernandez and Navarro~\cite{hernandez2012compressed,
hernandez2014compressed} propose to find dense subgraphs and represent $G$ as a
combination of (1) dense subgraphs $\mathcal{H}$ and (2) the rest of the graph
$\mathcal{R}$. For this, they first investigate the notion of dense subgraphs
and define dense subgraphs to be pairs $(S, C)$ of subsets of vertices, such
that each vertex belonging to $S$ is linked to each vertex belonging to $C$,
and simultaneously $S$ and $C$ are not necessarily disjoint. Thus, the pair
$(S, C)$ where $S = C$ indicates a clique and the pair $(S, C)$ where $S \cap C
= \emptyset$ indicates a biclique. The authors show that subgraphs where $S
\neq C$ (without being disjoint) occur often in web graphs and social networks,
and designing a compressed representation that utilizes the notion of dense
subgraphs pays off. Second, they store $\mathcal{H}$ using a combination of
integer sequences and bitmaps. To store $\mathcal{R}$, they use several
existing techniques such as $k^2$ trees~\cite{brisaboa2009k2,
brisaboa2014compact, ladra2011algorithms}, WebGraph schemes with the Layered
Label Propagation ordering~\cite{boldi2004webgraph, boldi2011layered}, and
$k^2$ partitioning~\cite{claude2011practical}.

Maneth and Peternek~\cite{maneth2016compressing, maneth2017grammar} recently
proposed a scheme that recursively detects substructures occurring multiple times in the same graph and
uses grammar rules to represent them. Moreover, they show that it is possible for some queries 
(e.g., reachability between two nodes) to be resolved in linear time when they are executed over the
grammar, enabling speedups proportional to the compression ratio.  The key
idea, similarly to Claude and Navarro~\cite{claude2007fast}, is to use
Re-Pair~\cite{larsson2000off}: \emph{``a phrase-based compressor that permits fast and
local decompression''}~\cite{larsson2000off}. However, contrarily to Navarro's work, they do not use
Re-Pair over a string built from the adjacency list, but instead represent
frequent substructures with grammar rules.

\subsubsection{Others}
\label{sec:webgraphs_others}

Guillaume et al.~\cite{guillaume2002efficient} aim to efficiently encode large
sets of URLs using only widely available tools, namely
\texttt{gzip} and \texttt{bzip}~\cite{deutsch1996gzip}. They also aim to make
the mapping between URLs and their identifiers as fast as possible, and to
compute $N_{out,v}$ efficiently. To avoid decompressing the entire
list of URLs, they split the sequence to be compressed into blocks and compress
each block independently. They utilize gap encoding, focusing on
differences between a given vertex and each of its neighbors to derive the
\emph{length} of each link. They also observe that such link lengths follow
a power distribution. Each length of a link is represented in either a
Huffman code, 16-bit integer, or 32-bit integer according to its absolute
value.

Asano et al.~\cite{asano2003compact} encode integers which have a power
distribution with a generalization of the variable-length nybble code.  They
use Kraft's inequality~\cite{kraft1949device} about instantaneous codes to show
that, when a random variable $X$ has a probability function $f(X)$, the
instantaneous code which minimizes average codeword length when used to
represent $X$ is $\log f(x)$ bits long when encoding $x$. Thus, if $X$ follows
the power distribution with the exponent $-\alpha$, the instantaneous code
minimizing the average codeword length is the variable-length block code with
$\frac{1}{\alpha - 1}$-bit blocks.
Next, they show that, when each $\mathscr{A}_v$ is gap encoded, the first
numbers in each $\mathscr{A}_v$ and the accompanying increments follows power
distributions of different exponents. They use this to develop a new encoding
of the web graph.  Consider the $\mathscr{A}_v$ of any $v$.  Suppose $v$ has
out-degree $d$. Then $\mathscr{A}_v$ has one initial distance and $d - 1$
increments. Consecutive 1s in the list of increments are compressed using the
run-length encoding~\cite{pountain1987run}.
Finally, the initial distance, the increments, and the run-length codes are
represented in the variable-length block codes with 6-bit, 3-bit, and 1-bit
blocks, respectively.

The main idea due to Asano et al.~\cite{asano2008efficient} is to identify
identical blocks in the adjacency matrix $A$ and then represent $A$ with a
sequence of blocks combined with some metadata information on the block type and
others. Now, they propose to use six different types of such blocks that
correspond to different types of locality within each host in the input web
graph. For a thorough analysis, they provide a detailed classification and an
extensive discussion on the proposed locality (and thus block) types.
Inter-host links are treated as related to a special type of locality.  
% 
% They achieve 1.99 bits per link to compress a graph with $m = 3,216,152$ and $n
% = 325,557$, outperforming the scheme due to Boldi and Vigna~\cite{sth} with
% 2.84 bits per link to compress the same web graph.

Apostolico and Drovandi target both web graphs and more general
graphs~\cite{apostolico2009graph}. Instead of naming vertices based on the
lexicographical ordering of their URLs (and thus being tailored for web graphs
only), they conduct a breadth-first search traversal of the input graph and
assign an integer to each vertex according to the order in which it is visited.
This ensures significant storage reductions after gap encoding is applied.
%
% Second, after the traversal, edges that constitute the traversal tree are
% encoded in a single sequence. Third, the remaining edges are encoded using
% gap encoding, however, the gap is computed not always with respect to the
% direct predecessor in the adjacency list, but in such a way that the edges
% encoded form consecutive chunks or a care 
%
They also introduce a new class of $\pi$-codes: universal codes for integers
that follow power law distribution with an exponent close to one.

Dhulipala et al.~\cite{dhulipala2016compressing} extend the work due to
Chierichetti et al.~\cite{chierichetti2009compressing} and \emph{''show how it can be
employed for compression-friendly reordering of social networks and web
graphs''}~\cite{dhulipala2016compressing}. 
They first note that optimal compression-friendly relabeling of vertices is
NP-hard. They then focus on reducing the problem domain size. For this,
they recursively bisect the graph and, once the size of the partitions is small
enough, compute a selected (possibly optimal) reordering for each partition.
Finally, these partial results are combined to obtain the solution for the
whole graph.

Analysis of the impact of various coding schemes on the compressibility of link
graphs was done by Hannah et al.~\cite{hannah2008analysis}.
Breyer~\cite{breyer2002web} presents the MarkovPR
software that optimizes storing URLs in web graphs with a large trie and
hashtables that alleviate navigating in the trie.
Finally, W-tree~\cite{avila2016w} is a space-efficient representation for web
graphs optimized for external memory settings. \maciej{extend}

\subsection{Social Networks}
\label{sec:social_networks}

Several recent works aim to specifically condense social networks. Some offer
novel schemes while others investigate how to reuse the schemes developed for
web graphs.

Chierichetti et al.~\cite{chierichetti2009compressing} provide three
contributions targeted at social network compression. First, they prove
hardness results about several types of vertex reordering; we provide more
details in a section devoted to vertex relabeling (\cref{sec:relabeling}).
Second, they propose the BL compression scheme that extends Boldi and Vigna's
BV scheme from the WebGraph framework. BL takes
advantage of a certain property common in social networks, namely
\emph{reciprocity}, next to the properties of locality and similarity.
Reciprocity means that most unidirectional links are reciprocal, i.e., there is
a link with a reverse direction connecting the same vertices.  Third, the final
design contribution is the \emph{shingle} ordering that preserves both locality
and similarity.  Intuitively, it treats $N_{out,v}$ as a set and derives a special
value called the \emph{shingle} $M_\sigma(N_{out,v})$ of this set where
$\sigma$ is a suitably selected permutation (or hash function). Then, the
vertices of the input graph are ordered by their shingles. The authors show
that, if two vertices have many outneighbors in common, then the probability that 
they have an identical shingle (and thus are not far from each other when considering the shingle
ordering) is high.

% The nodes in V can then be ordered by the shingles. By the property
% stated above, if two nodes have significantly overlapping outneighbors,
% i.e., share a lot of common neighbors, then with high
% probability they will have the same shingle and hence be close to
% each other in a shingle-based ordering.
% 
% ``It is based
% on obtaining a fingerprint of the out-neighbors of a vertex and ordering
% the vertices according to this fingerprint.''
% To explain it in more detail, recall the definition of the \emph{Jaccard coefficient}
% $J(A,B)$ of sets $A, B$: $J(A,B) = |A \cap B| / |A \cup B|$; it is a well-known
% measure of the similarity of sets $A,B$.

Maserrat and Pei~\cite{maserrat2010neighbor, maserrat2012compression} aim to answer both $N_{in,v}$ and $N_{out,v}$ in
sublinear time in $n$ and $m$ while compressing the input graph. 
For this, they propose an \emph{Eulerian data
structure}: a structure that stores a linearization of the input graph
in a space-efficient way and uses it to answer the neighborhood queries efficiently.
\maciej{TODO: possibly extend.}

Boldi et al.~\cite{boldi2011layered} propose Layered Label Propagation (LLP), a
compression-friendly vertex ordering targeting social networks. They start
their work with an analysis that aims to formally understand why the existing
approaches compress web graphs well (see~\cref{sec:webgraphs_others} for a
detailed discussion on this part of their work). Still, the bulk of the paper
is dedicated to the LLP ordering that targets social networks in the first
place.
To understand LLP, we first explain three other related schemes: a generic
label propagation algorithm, a simple label propagation algorithm
(LPA)~\cite{raghavan2007near} and a variant of the Absolute Potts Model (APM)
scheme~\cite{ronhovde2010local} that builds upon LPA.

First, any label propagation algorithm executes in rounds. During every round, 
the label of each vertex is updated; the exact form of an update is determined
by a pre-specified rule. 
Different rules constitute the difference between different label propagation algorithms.
Before the first iteration each vertex has a different label; the algorithm
terminates when no more update takes place. 

Second, in LPA, a vertex decides to adopt a label that is used by most of its
neighbors. Its main problem is that it often generates a single giant cluster with
most of the vertices.

APM addresses this issue. Assume a vertex $v$ has $k$ neighbors and let
$\lambda_1, ..., \lambda_k$ be the labels belonging to $v$'s neighbors. Let
also $k_i$ and $v_i$ be the number of $v$'s neighbors with a label~$\lambda_i$ and
the total number of vertices in $G$ with $\lambda_i$, respectively.
Now, when updating its label, instead of selecting a label~$\lambda_i$ that has
the maximum value of $k_i$, $v$ selects a label that maximizes the value
$k_i - \gamma (v_i - k_i)$.  Intuitively, this rule does not only increase the
density (i.e., the number of edges) of a given community (which happens because
$k_i$ new edges adjacent to $v$ join a given community), but also decreases
it because of $v_i - k_i$ non-existing edges. The $\gamma$ parameter controls
the importance of each of these two effects. This strategy prevents
generating one huge giant cluster.

Now, to understand the idea behind the LLP scheme, first observe that different
values of $\gamma$ unveil clusters of different resolutions. If $\gamma$ is
close to 0 it highlights large clusters (when $\gamma = 0$ then APM degenerates
to LPA); increasing $\gamma$ unveils small clusters.  LLP attempts to obtain a
labeling that considers clusters of various resolutions.  In general, it
iteratively executes APM with various $\gamma$ values. Now, each such iteration
outputs a vertex labeling.  Vertices within the same cluster maintain the same
order from past iteration.  Vertices that acquired the same label are attempted
to be placed as close to one another as possible.

% Moreover, Boldi et al.~\cite{boldi2009permuting_social} test various existing
% vertex labelings and propose new ones for both social and web graphs. Among
% others, they illustrate that the URL ordering is much less effective on
% transposed web graphs.  Then, they show that their labelings designed for web
% graphs such as the Gray code ordering also offer promising results for social
% networks, being comparable in storage reduction to the orderings due to
% Chierichetti et al.~\cite{chierichetti2009compressing} while offering fast
% access to the compressed data.

Boldi et al.~\cite{boldi2010compressing} show in more detail advantages of
using a clustering algorithm described in the LLP paper~\cite{boldi2011layered}
for web graphs. They discuss how to use it to enhance both locality and
similarity and provide several interesting examples visualized with
associated adjacency matrices.

Shi et al~\cite{shi2012optimizing} illustrate that the $k^2$ tree
representation (\cref{sec:k2_web}) can be enhanced in several ways.  Among
others, they propose to use the DFS vertex order combined with a heuristic that reorders the adjacency matrix to make sure the cells with ``1''
are concentrated in few submatrices. To acheive this, the heuristic huses the Jaccard coefficient for the structural similarity of any two
vertices.
% 
% Second, they use the kcto indicate if a social network achieved
% \emph{maturity} (i.e., the overall behavior of the network is homogeneous
% even if individual vertices may express heterogeneous types of behavior).

Liakos et al.~\cite{liakos2014pushing, liakos2014effect} use the fact that the
LLP reordering enhances the locality in such a way that the corresponding AM
contains a large ``stripe'' around its diagonal that groups a large fraction of
edges. They use a bitvector to represent these edges and ultimately reduce
space to store a network.
Finally, Cohen briefly discusses various strategies for social network
compression~\cite{cohen2016data}.

\subsubsection{Combining Web and Social Networks}

Among the works described in~\cref{sec:web_graphs}
and~\cref{sec:social_networks}, some are dedicated to compressing both web and
social networks~\cite{chierichetti2009compressing, maserrat2010neighbor,
dhulipala2016compressing, boldi2009permuting_social, boldi2010compressing,
claude2011practical, hernandez2012compressed, hernandez2014compressed,
hernandez2011compression_web_social}.
Zhang et al.~\cite{zhang2014common} propose the \emph{bound-triangulation}
algorithm.  The main idea is to use a data structure that stores triangles
efficiently. The motivation is that many web graphs and social networks
contain a large number of triangles, thus priority placed over storing this
motif efficiently reduces the required storage.
Angelino~\cite{angelinocompressing} proposes a new vertex ordering that
considers semantic data associated with the graph. For example, they propose to
sort vertex neighborhoods by a selected property such as ``name''.
Miao~\cite{miao2016experimental} extracts dense subgraphs from web and social
graphs and encodes them using succinct data structures such as wavelet trees.

\subsection{Biological Networks}

A significant amount of work is dedicated to compressing biological networks. The
vast majority are related to genome assembly
networks~\cite{sheikhizadeh2016pantools}. Besides that, few others exist, for
example on compressing gene regulatory networks~\cite{dinkla2012compressed} and
metabolic graphs~\cite{ay2012metabolic}, or optimizing protein network
alignment~\cite{kamal2014memory}. There also exists a
survey~\cite{hosseini2016survey} on compressing various types of biological
data (not necessarily graphs). 

% \subsubsection{Related Concepts}
% 
% We first survey the concepts used heavily in this section.
% 
% \macb{}

\subsubsection{Schemes Based on De Bruijn Graphs}
\label{sec:de_bruijn_biology}

De Bruijn graph~\cite{sainte1894solution, good1946normal, de1946combinatorial}
is a directed graph that represents overlaps between sequences of symbols. For
a given set of symbols of cardinality $s$, the corresponding $N$-dimensional De
Bruijn graph contains $s^N$ vertices; these vertices consist of all possible sequences
of these symbols; each symbol may appear more than once in a sequence. An
edge from a vertex $v_1$ to a vertex $v_2$ exists iff we can shift all the
symbols associated with $v_1$ by one position to ``the left'' and fill the free position at the end with another symbol, to ultimately
obtain the sequence associated with $v_2$.

De Bruijn graphs are commonly used in the \emph{de novo}
genome assembly~\cite{bankevich2012spades, butler2008allpaths, li2010novo, peng2010idba,
compeau2011apply, zerbino2009genome, flick2015parallel, sohn2016present,
crusoe2015khmer}, which is one of fundamental bioinformatics projects. 
Some specific applications include assembly of DNA
sequences~\cite{idury1995new}, mRNA~\cite{grabherr2011full} assembly,
metagenome assembly~\cite{peng2011meta}, detection of genomic variants~\cite{peterlongo2010identifying, iqbal2012novo} and calling of \emph{de novo}
alternative splicing~\cite{sacomoto2012k}.

In the process of genome assembly, long contiguous DNA sequences are constructed
from many much shorter DNA fragments The long parts are called contigs while the short parts are called reads. Assemblers
based on De Bruijn graphs first extract subsequences (mers) of length $K$
from reads ($K$ is a parameter). Then, a De Bruijn graph consisting of mers
as vertices is built and then simplified, if possible. Now, contigs are simple
paths in this graph and then can be extracted by finding a Hamiltonian or (more
preferably) an Eulerian path.

A serious space and time bottleneck is the construction and navigation of the
graph.  This is why space-efficient representations of de Bruijn graphs have
been researched intensely. The storage lower bound ([bits]) of a De Bruijn
graph constructed from $K$-mers is $\log_2 \binom{4^{K+1}}{m}$ ($m = |E|$).

% \begin{gather}
% %
% \log_2 \dbinom{4^{K+1}}{m} 
% %
% \end{gather}

Li et al.~\cite{li2010novo} were the first to use De Bruijn graphs in assembly
of human genome with mers large enough to detect structural variation between
human individuals, to annotate genes, and to analyze genomes of novel species.
They used minimum-information de Bruijn graphs without the information on read
locations and paired-end information.
There were various previous short-read assemblers, including
EULER~\cite{pevzner2001eulerian}, Velvet~\cite{zerbino2008velvet},
ALLPATHS~\cite{butler2008allpaths}, and EULER-SR~\cite{chaisson2008short}.
Yet, they all are targeted at bacteria- or fungi-sized genomes, and are mostly
unable to manage large genomes. ABySS is another assembler, implemented with
MPI on distributed-memory machines for more
performance~\cite{simpson2009abyss}. It avoids using pointers in a De Bruijn
graph representation for memory savings; the graph is represented as a
distributed hash table, acting as a mapping from a $K$-mer to a byte with the
connectivity information related to this mer.

Ye et al.~\cite{ye2012exploiting} show how to construct a graph equivalent to
the de Bruijn graph while maintaining only one (out of $g$) vertices ($g \in [10;25]$).
They skip a fraction of $K$-mers to reduce memory
consumption. As an example, assume that there are two pairs of overlapping
vertices: $A,B$ and $B,C$. The authors then store the $(A, C)$ overlap; two overlaps $(A,B)$ and $(B,C)$
are not stored, eliminating read $B$ from the
graph. They attempt to sample one out of every $g$ ($g < K$) $K$-mers.

Cazaux et al.~\cite{cazaux2016linking}, Minkin et al.~\cite{minkin2016twopaco},
and others~\cite{beller2015efficient, georganas2014parallel,
rengasamy2017parallel, limon2014efficient, rozov2017faucet} show fast and
space-efficient algorithms for constructing compact De Bruijn graphs.

Other works include building a space- and time-efficient index used for
pattern-matching in De Bruijn graphs~\cite{almodaresi2017space} and compacting
De Bruijn graphs with little memory~\cite{chikhi2016compacting}.
Various other works exist~\cite{marcus2014splitmem, lin2014manifold, peng2011t,
goldstein2004sequences, jaillard2017representing, cazaux2014indexing,
beller2016representation, almutairy2013space, chikhi2014representation}.  Finally, for completeness, we also mention studies
into probabilistic De Bruijn graphs~\cite{pell2012scaling,
benoit2014compression, benoit2015reference}.

% Beller and Ohlebusch~\cite{beller2016representation} propose a new
% space-efficient representation of the compressed de Bruijn graph with an
% option to search for a pattern (e.g. a specified  variant form of a gene)
% within the pan-genome.

Current main approaches for a compact De Bruijn graph representation as their
foundation use
Bloom filters~\cite{bloom1970space}, Burrows-Wheeler
Transform~\cite{burrows1994block}, and succinct data structures
(\cref{sec:bounds}).

\macb{Schemes Based on Bloom Filters}
Chikhi and Rizk~\cite{chikhi2013space} use a Bloom filter to maintain edges (with
additional structures to avoid false positive edges that would affect the
assembly). 
% 
% They traverse the graph by generating all possible outgoing edges at
% each vertex and testing their membership in the Bloom filter.
%
Next, Salikhov et al.~\cite{salikhov2013using} design \emph{cascading} Bloom
filters to outperform storage requirements of Chikhi and Rizk's approach.  They
change the representation of the set of false positives. The key idea is to
\emph{``iteratively applying a Bloom filter to represent the set of false
positives, then the set of ``false false positives'',
etc.''}~\cite{salikhov2013using}. This cascade enables 30\% to 40\% less memory
with respect to Chikhi and Rizk's method~\cite{salikhov2013using}. 
Other authors used Bloom filters to implement de Bruijn graphs for
pan-genomics~\cite{holley2015bloom} and to enhance connecting
reads~\cite{vandervalk2014konnector}.
A redesign of the ABySS scheme was recently implemented using Bloom
filters~\cite{jackman2017abyss}.

\macb{Schemes Based on Succinct Data Structures}
Conway and Bromage~\cite{conway2011succinct} use succinct (entropy-compressed,
see~\cref{sec:bounds}) data structures for a \emph{``practical representation of the De
Bruijn assembly graph''}~\cite{conway2011succinct}.  They
use~\cite{Okanohara07practicalentropycompressed} as succinct representations of
a bitmap used to represent De Bruijn graphs.
Bowe et al.~\cite{bowe2012succinct} also incorporate succinctness.  They show a
representation that uses $4m+o(m)$ bits of a De Bruijn graph with $m$ edges and
ensure various graph queries in constant time (for example, computing the in-
and out-degree of a vertex).  The structure is constructed in $O(N K \log m /
\log \log m)$ time using no additional space where $K$ and $N$ and lengths of
mers and the whole DNA, respectively.
The authors combine (1) succinct static
strings due to Ferragina et al.~\cite{ferragina2007compressed}, (2) succinct
dynamic strings~\cite{navarro2014fully}, and (3) the XBW-transform
structure~\cite{ferragina2009compressing}.
Bowe et al.'s work was expanded by Boucher et
al.~\cite{boucher2015variable}, Belazzougui et
al.~\cite{belazzougui2016bidirectional,belazzougui2016fully}, and Pandey et al.~\cite{pandey2017debgr}.
Succinct colored De Bruijn graphs were also discussed~\cite{belk2016succinct,
almodaresi2017rainbowfish}.

\macb{Schemes Based on Burrows-Wheeler Transform}
Various works incorporate the Burrows-Wheeler Transform for more space
efficiency~\cite{baier2015graphical, liu2016debwt, giessepan,
rodland2013compact}

\subsubsection{Grammar- and Text-Related Works}
Peshkin~\cite{peshkin2007structure} uses the notions from both graph grammars
and graph compression to understand the structure of DNA and simultaneously be
able to represent it compactly. He proposes the Graphitour
algorithm that finds a simplified graph to
construct the input graph representing the structure of a given DNA sequence.
The simplification scheme is based on contracting edges that satisfy certain
criteria regarding their similarity.
Next, Hayashida and Akutsu~\cite{hayashida2010comparing} use and extend Graphitour to
be able to compare two different biological networks. Specifically, they assess
the similarity of two networks by comparing the compression ratios of these two
networks when compressed using the modified Graphitour variant. The work is
applied to various metabolic networks.
Finally, there are other grammar- and text-related works~\cite{goldstein2004sequences}
that treat genome sequence as piece of text.

\subsubsection{Hierarchical Approaches}

Hierarchical approaches based on merging groups of vertices into supervertices
can also be found in this domain.  Brown et al.~\cite{brown2016evolving}
consider two vertices similar if a high proportion of their neighbours are
common.  Such vertices are merged to form supervertices.
Other similar approaches uses genetic algorithms to find similar vertices
efficiently~\cite{collins2017single, zakirov2017nsga}.

\subsubsection{Others}

Other approaches include novel types of space-efficient graphs such as
Superstring graphs~\cite{cazaux2016superstring} or compressing frequent motifs
in a given biological network for not only storage reductions but also faster
discovery of various patterns~\cite{wang2012symmetry}.

\subsection{RDF Graphs}
\label{sec:rdf_graphs}

The Resource Description Framework (RDF) is a set of World Wide Web Consortium
(W3C) specifications that provide semantic information in a format
interpretable by machines. An RDF graph can be modeled as a set of triples
consisting of a \emph{subject}, a \emph{predicate}, and an \emph{object}. Any
of the triple elements can be a string; storing the triples explicitly can be
memory intensive. One can thus assign identifiers to such values and use a
dictionary to map them to concrete value. Consequently, to compress RDF graphs
one can compress the dictionary or the underlying graph structure.

\subsubsection{Modeling RDF Graphs As Relational Databases}

Early approaches for compressing RDF graphs map the graphs to relational
databases. One way is to simply store all RDF triples in a \emph{triple store}:
a table with 3 attributes (columns), an approach used in RDF storage systems
such as Jena~\cite{mcbride2001jena, khadilkar2012jena},
Sesame~\cite{broekstra2002sesame}, and 3store~\cite{harris20033store}. 
Another approach is to use \emph{property tables}.  In an example scheme,
several tables can be built and the attributes in each are properties shared by
the triples; the remaining triples that do not fit into the property tables are
stored in a triple store~\cite{wilkinson2006jena}.
Third, researchers also proposed vertical
partitioning~\cite{abadi2007scalable}, a scheme where there is one table per
one property. The core idea is thus to group triples by predicate, generating
many 2-attribute tables (one for a single predicate value).
Finally, other works include space reduction schemes in the
Hexastore~\cite{weiss2008hexastore}, RDF-3X~\cite{neumann2010rdf},
TripleBit~\cite{yuan2013triplebit}, or BitMap~\cite{atre2010matrix} systems.
RDF-3X and BitMap use gap compression in various parts of the system; for
example, RDF-3X condenses indexes in leaves of the underlying
B+-tree~\cite{abel1984b+}.

\subsubsection{Understanding and Utilizing RDF Redundancy}

Pan et al.~\cite{pan2014ssp} first categorize the redundancy in RDF graphs into
three different types: \emph{semantic}, \emph{syntactic}, and \emph{symbolic}.
Semantic redundancy can be found in RDF graphs that use more triples than
necessary to describe a given set of data (i.e., they are not semantically
richer than their subgraphs with fewer triples).  Syntactic redundancy can be
found in graphs that use excessive syntax (e.g., a plain list of triples)
instead of a more compact one (e.g., binary serialization).  Finally, symbolic
redundancy takes place when the average bit count required 
for encoding a basic symbol (an RDF resource) is not optimal. After the
  redundancy analysis, the authors propose to compress RDF datasets by using
  frequent graph patterns to remove all three aforementioned types of
  redundancies.

In another piece of work, Pan et al.~\cite{pan2014graph}
exploit the RDF graph structure to enhance compression at both the semantic and
syntactic level. For the semantic level, they develop a generic framework to
replace instances of the bigger graph patterns with smaller instances of the
smaller graph patterns (i.e., they eliminate semantic redundancies).  This
approach is similar to the grammar-based compression schemes for web graphs
where more complex generation rules were replaced with simpler and smaller
ones.  Moreover, for the syntactic level, they illustrate that the same set of
RDF triples can occupy various amounts of space depending on how triples are
serialised in an RDF file.  They identify \emph{intra-structure redundancies}
(multiple occurrences of identical RDF resources within the same RDF subgraph)
and \emph{inter-structure redundancies} (multiple occurrences of identical
resources across different RDF subgraphs).

Moreover, Fernandez et al.~\cite{fernandez2010rdf} analyzes the compressibility
of RDF data sets. Specifically, the authors show that large RDF graphs can
effectively be compressed because of the power law vertex degree distribution,
the hierarchical organization of URLs, and the verbosity of the RDF syntax.
Esposito et al.~\cite{esposito2005redd} develop algorithms for detecting
various RDF redundancies.
Pichler et al.~\cite{pichler2013complexity} analyze the complexity of detecting
redundancy in RDF datasets.
Wu et al.~\cite{wu2014redundant} categorize RDF redundancy redundancy and
design new methods for detecting redundancy.
Finally, various schemes and analyses of redundancy elimination in RDF graphs were
proposed and conducted by Pichler et al.~\cite{pichler2010minimising,
pichler2010redundancy}, Meier~\cite{meier2008towards}, Iannone et
al.~\cite{iannone2005optimizing}, Grimm and
Wissmann~\cite{grimm2011elimination}, 

\subsubsection{Incorporating HDT Structure}

Fernandez et al.~\cite{fernandez2013binary, fernandez2011lightweighting} design
Header-Dictionary-Triples (HDT): an RDF representation that partitions RDF
graph data into three modules dedicated to \emph{``the header information, a
dictionary, and the actual triples' structure''}~\cite{fernandez2013binary}.
The modular design reduces redundancy and limits required storage by up to more
than an order of magnitude. HDT takes advantage of the fact that the items in
the considered RDF datasets follow the power law distribution. 
% 
% Moreover, the main
% factor in reducing the size is not
% reduction is achieved due to a more condensed representation rather than a
% reduction in the number of triples.
%
Next, Fernandez et al.~\cite{fernandez2014efficient} compress RDF streams by
proposing the Efficient RDF Interchange (ERI) format that \emph{``exploits the
regularity of RDF streams''}~\cite{fernandez2014efficient}.
Hernandez-Illerai et al.~\cite{hernandez2015serializing} extend HDT with HDT++. 
HDT++ alleviates various redundancies (e.g.,
they group objects per predicate). They ultimately compress some popular RDF
datasets by more than 50\% and outperform the state-of-the-art $k^2$ trees in
size by 10–13\%.

\subsubsection{Incorporating MapReduce}

Several works attempt to reduce the size of RDF datasets with MapReduce
(MR)~\cite{dean2008mapreduce}.  Gimenez-Garcia et al.~\cite{gimenez2015hdt} use
MR to process large RDF datasets and serialize them into the
Header-Dictionary-Triples (HDT) format~\cite{fernandez2013binary} that reduces
storage overheads behind RDF graphs.
Urbani et al.~\cite{urbani2013scalable} propose to use MR to overcome the
scalability problems of compressing large RDF graphs. Specifically, they use MR
to construct an RDF dictionary.
Similarly, Cheng et al.~\cite{cheng2014efficient} also reduce the size of RDF
graphs; they use the X10 language~\cite{charles2005x10} to construct RDF
dictionaries.
There are other similar approaches~\cite{huang2011scalable, urbani2010massive,
gimenez2015scalable}.

\subsubsection{Generating Equivalent and Smaller Rules}

Joshi et al.~\cite{joshi2013logical, joshi2017exploiting, joshi2012towards}
propose Rule Based Compression (RBC): a compression technique for RDF datasets
that \emph{``compresses datasets by generating a set of new logical rules from
the dataset and removing triples that can be inferred from these
rules''}~\cite{joshi2013logical}.  The authors show that RBC can
remove up to half of the initial triples while maintaining integrity of data.
%
% The core idea is to infer triples from several logical rules associated with
% a given dataset.  
%
For example, a triple $<A, \text{grandfather--of}, C>$ can be generated from triples
$<A, \text{father--of}, B>$, $<B, \text{father--of}, C>$ assuming the introduction of an ontology
appropriately connecting relations $\text{father--of}$ and $\text{grandfather--of}$.

Fernandez et al.~\cite{fernandez2014rdsz} propose a scheme called RDF
Differential Stream.  It uses structural similarities among items in a stream
of RDF triples and combines differential encoding with zlib.
Zhang et al.~\cite{zhang2014assg} compress RDF datasets with Adaptive
Structural Summary for RDF Graph (ASSG): a compression method that uses
bisimulation~\cite{dovier2004efficient} to create an equivalent graph of
smaller size where vertices with identical labels are collapsed into fewer
vertices.
Lyko et al.~\cite{lykoscaro} use logical implications contained in the data to
develop rules and simultaneously minimize the number of triples that need to be
stored.
Gayathri et al.~\cite{gayathri2015horn} mine logical Horn
rules~\cite{horn1951sentences} in RDF graphs and then store only the
triples \emph{``matching the antecedent part of the rules''}~\cite{gayathri2015horn}. Triples matching the
head part of the rules can be generated by applying these
rules and they are thus deleted.
Guang et al.~\cite{guang2016detect} propose rule-based methods to find and
delete semantically redundant triples.

\subsubsection{Incorporating Hierarchical Schemes}

Fernandez et al.~\cite{fernandez2010compact} compress RDF graphs by grouping
triples with the common subject into adjacency lists. Then, for each RDF
property value and subject, it stores ordered IDs of the associated objects.
The derived ID sequences are treated with Huffman encoding and PPMd
7-zip~\cite{pavlov20077zip}.
%
% Another scheme that ensures both good compression and high performance
% queries is based on the $k^2$-tree~\cite{brisaboa2009k2} data structure.  The
% core idea is to maintain a single $k^2$-tree for a single RDF property.
%
Next, Jiang et al.~\cite{jiang2013graph} propose two schemes. First, they assign a
\emph{type} to each RDF object and subject and then reduce the number of
vertices in the RDF graph by grouping and collapsing RDF entities with the same
type. Second, they compress the RDF graph by removing vertices with only one
neighbor and maintaining the information on the removed vertex at its neighbor.
Finally, Bazoobandi et al.~\cite{bazoobandi2015compact} develop a new
Trie structure variant~\cite{de1959file} and use it as a dictionary for RDF datasets in
a dynamic and streaming setting. They specifically alleviate common prefixes
found in strings in RDF datasets.

\subsubsection{Compressing RDF Dictionaries}

Martinez et al.~\cite{martinez2012compression, martinez2012querying}
specifically compress RDF dictionaries. Among others, they apply
existing techniques for compressing string dictionaries, including a compact
form of \emph{hashing}~\cite{cormen2009introduction, brisaboa2011compressed},
\emph{Front-Coding}~\cite{witten1999managing} (both Plain
Front-Coding~\cite{brisaboa2011compressed} and Hu-Tucker
Front-Coding~\cite{knuth1998art}), and various forms of
\emph{self-indexing}~\cite{navarro2007compressed_fti,
ferragina2000opportunistic, burrows1994block, brisaboa2011compressed}.
Moreover, Dawelbeit and McCrindle~\cite{dawelbeit2016efficient} compress RDF dictionaries
that are used in Google BigQuery.

\subsubsection{Using $k^2$ Trees}

Another recent work~\cite{alvarez2011compressed, alvarez2013compressed,
alvarez2015compressed} combines vertical partitioning~\cite{abadi2007scalable}
with $k^2$-trees~\cite{brisaboa2009k2, brisaboa2014compact}. The core
technique is called $k^2$-triples. It first vertically partitions the dataset
into subsets of pairs (subject, object); these subsets are disjoint.
%
% one subset per predicate.
%
Next, they use binary matrices to represent these subsets of pairs where one cell
with ``1'' indicates that a given triple exists in a given RDF graph. These
matrices turn out sparse and they are then encoded with $k^2$-trees.
A related work by Alvarez et al.~\cite{garcia2014interleaved,
alvarez2017succinct} advocates \emph{Interleaved} $k^2$ trees: a compressed and
self-indexed representation with efficient
querying of general ternary relations, similar to $k^2$ trees and their application in compressing 
binary relations.  The main idea is to represent a given set of
triples as $x$ binary relations and then use $x$ $k^2$ trees 
and gather them within a single tree. Interleaved $k^2$
trees can be applied to generic ternary relations; they are evaluated on RDF.
Next, Brisaboa et al.~\cite{brisaboa2017compressed} propose a dynamic variant
of the $k^2$ tree data structure to compactly represent binary relations.
The above-described efforts into combining RDF datasets and
$k^2$ trees were also described in theses by
Alvarez~\cite{alvarez2014compact} and Roca~\cite{de2014new}.

\subsubsection{Using Succinct Data Structures}

Cure et al.~\cite{cure2014waterfowl} use succinct data structures to compress
RDF data and to ultimately alleviate scalability issues.

\subsubsection{Others}

Other efforts include the following related work.
Swacha and Grabowski~\cite{swacha2015ofr} compress RDF datasets with a
combination of techniques. They separate semantic (i.e., RDF specialized) and
general-purpose encoding. They also separate graph and dictionary compression,
and combine various techniques over the contents (e.g., run-length encoding or
reordering the content).
Zneika et al.~\cite{zneika2015rdf} summarize RDF datasets by adding the
information on various instances of data patterns for more performance.
Fernandez~\cite{fernandez2011compact} compacts RDF datasets with a combination
of various techniques.
Jagalpure~\cite{jagalpure2012rgis} designs novel indexing techniques for more scalable and
storage-efficient RDF databases.
Weaver and Williams incorporate a subset of the Turtle
syntax~\cite{beckett2008turtle} and Lempel-Ziv-Oberhumer (LZO) compression to
reduce I/O load in parallel RDF systems.
Joshi et al.~\cite{joshi2015alignment} exploit ontology alignments and
application context in RDF graphs for compression.
Gallego et al.~\cite{gallego2013compressing} focus on compressing RDF data in
the context of multimedia retrieval.
%
%
% Pappas et al.~\cite{pappas2017exploring} discuss summarizing various
% centrality graph measures in the RDF setting.
%
Deme et al.~\cite{deme2018efficient} enhance the design of the RDSZ scheme.
Brisaboa et al.~\cite{brisaboa2015compact} propose a novel RDF storage scheme
called RDFCSA that combines the data and the associated index in a single
representation and builds on suffix arrays.
Bit vectors for compressing RDF are used by Atre~\cite{atre2011bit}.
Various other RDF compression schemes exist~\cite{cure2015litemat, urbani2016kognac, gayathri2015horn, lhez2017compressed, kabisch2015standardized, fernandez2016evaluating, fernandezhdtq, reindorfhdt}.

\subsection{Network Graphs} 

Some works target compressing graphs originating in the area of networking.
%
% Gilbert and Levchenko~\cite{gilbert2004compressing} consider creating graphical
% summaries of enterprise IP networks in a process that they call \emph{semantic
% graph compression}.  Their goal is not only to reduce the size of graphs but
% also create graph summaries that can be viewed and analyzed graphically by
% humans.  They use two schemes for this: (1) calculating the importance of each
% vertex according to a selected importance measure and preserving only the
% ``important'' vertices, and (2) obtaining the similarity measure of different
% vertices and then combining similar vertices into a single vertex.
% 
Gilbert el al.~\cite{gilbert2004compressing} summarize IP networks to
facilitate visualization. They refer to it as the \emph{``\emph{semantic} graph
compression to distinguish it from the \emph{algorithmic} graph compression
where a graph is compressed in order to reduce the time or space complexity of
a graph algorithm''}~\cite{gilbert2004compressing}. They preserve selected
properties, for example connectivity: the compressed graph should be connected
if the original graph is connected. Moreover, they also develop compression
  schemes that collapse similar vertices into one (hierarchical schemes). The
  similarity measure only uses information about graph topology (i.e.,
  structure of connections between vertices) or vertex or edge properties. 

Jusko et al.~\cite{jusko2014memory} use Bloom filters~\cite{bloom1970space} to
develop representations of connection graphs (e.g., P2P overlays) that reduce
the amount of consumed memory; the targeted setting is Software Defined
Networking (SDN)~\cite{kreutz2015software}. The representation enables network
elements to determine which connections are to be escalated for further
processing and it simultaneously prohibits extracting any other information
from the graph for security. The connection graph is assumed to be dynamic.

Shi et al.\cite{shi2013scalable} compress network traffic graphs by grouping
motifs such as a clique into single vertices; thus again incorporating
a hierarchical compression scheme.

Other works include compressing changes in network monitoring
data~\cite{cheng2014exploiting}.

\subsection{Chemistry Networks}

There are only very few works related to compressing graphs used in various
chemical sciences. In general, there exist some studies on applying graph
theory in chemistry~\cite{balaban1985applications}.  Compression of such graphs
was mostly not addressed.  Burger et al.~\cite{burger2016compressed,
burger2017memory} address compressing graphs used to model Super Carbon
Nanotubes (SCNTs).
Example such graphs are Hierarchically Symmetric Graphs
(HSG)~\cite{schroppel2015meshing}. They can model hierarchical structure of
SCNTs.  The authors present the Compressed Symmetric Graphs (CSG) that is
constructed out of the description of an HSG while exploiting the structural
symmetry in the HSG to only store nodes and edges required for efficiently
reconstructing requested parts of the original graph on-the-fly.

\subsection{Geographical Datasets}

% Compressing various graph-related datasets in geographical sciences is also
% not an extensive topic. On general, one may distinguish two subareas.

One may distinguish two subareas
in compressing graph-related datasets in geographical sciences.

\subsubsection{Compressing Terrain Datasets}

Geography Information System (GIS) data is usually 3D terrain data.  Thus, is
it stored with 3D meshes. For compressing such datasets, one could use any of
the available generic schemes for mesh compression; they were covered in
several surveys~\cite{maglo20153d, peng2005technologies,
szymczak2002piecewise}. There are also works related to specifically
compressing GIS data. For example, Pradhan et al.~\cite{pradhan2007second}
target GIS data with Delaunay triangulation~\cite{george1998delaunay}.

\subsubsection{Compressing Raster Datasets}

This subdomain is mildly related to graph compression, we still present it for
completeness. In short, \emph{''raster data (...) is commonly used [in GIS] to represent
attributes of the space (temperatures, pressure, elevation measures, etc.)''}~\cite{ladra2016compact}. Now, these
sets can often be represented as matrices.  Ladra et
al.~\cite{ladra2016compact, ladra2017scalable} use two ideas to compress such
matrices: they first construct a special structure called $k^2$ raster that is
based on $k^2$ trees.  On top of $k^2$ trees, the nodes in the tree maintain
the maximum and minimum values of each submatrix. These matrices are used to represent the
raster data. This also provides the indexing
functionality.  Selected compact data structures are then used to encode
elements of $k^2$ raster, such as the tree structure.

\subsection{VLSI Graphs}

Yang et al.~\cite{yang2008approach} compress VLSI structures. They focus on the
EDIF (Electronic Data Interchange Format)~\cite{edif1987edif} data format.
Their main idea is to use various data mining algorithms to discover
redundancies, for example multiple identical subgraphs, and use the redundancies for
compression.

\section{Compressing Graph Databases}
\label{sec:databases}

We now consider works dedicated to compressing graph databases.  Graph database
is a database that stores graphs and enables semantic queries over them.  An
example of such a database is neo4j~\cite{webber2012programmatic} or
G*~\cite{labouseur2015g} that specifically targets compressing dynamic graphs.

\subsection{Bitmap-Based Schemes}

DEX~\cite{martinez2011dex} is a general-purpose system for managing and
processing large graphs. It uses an internal representation 
based on compressed bitmaps~\cite{martinez2012efficient} for efficient
basic navigation operations.

\subsection{$k^2$ Tree-Based Schemes}

Lehmann and Perez~\cite{lehmann2015implementing} report empirical results on
implementing graph queries over graph representations compressed with $k^2$
trees. They focus on two-way regular-path queries (2RPQs) as these queries can
express navigating graphs with paths defined by regular expressions.
Alvarez et al.~\cite{alvarez2010compact} present a new model and representation
of general graph databases, where nodes and edges are typed, labeled, and
possibly attributed, and graphs may be multigraphs.  They also discuss
efficient implementation of graph navigation operations.
Specifically, they propose the Compact Graph Database structure in which
any multigraph is represented using three $k^2$ trees for three
relations: (1) a relation between nodes and their attributes, (2) a relation
between edges and their attributes, and (3) a relation between nodes (i.e., the
actual edges).
An interesting aspect of their work is a formal model of a labeled, attributed,
and typed multigraph, that is a 10-tuple $(\Sigma_N, \Sigma_E, N, E, ST,
\Sigma_A, NS, ES, NA, EA)$.  $\Sigma_N, \Sigma_E$ are sets with node and
edge types; $N, E$ are sets of pairs that associate numeric node or edge
numeric identifiers with their types; $ST$ is a set of pairs that associate an
edge numeric identifier with a pair of this edge source and destination nodes;
$\Sigma_A$ is a set containing attribute names; $NS, ES$ are schemes that
describe attributes of each node or edge type; $NA, EA$ are sets with pairs
associating node or edge attributes and their values.

\vspace{-0.25em}
\subsection{Succinct Data Structures}

In the ZipG system~\cite{khandelwal2017zipg}, the input graph data is transformed
into two flat unstructured files that contain the vertex and edge info,
respectively.  In addition, these files also store some (small) metadata amount
to enable efficient interactive queries.
Now, the ZipG is implemented on top of Succinct~\cite{agarwal2015succinct}, a
distributed data store that compresses unstructured data and key-value pairs
and offers
random access and arbitrary
substring search queries.

\vspace{-0.25em}
\subsection{Hierarchical Schemes}

Maccioni and Abadi~\cite{maccionicompressing} introduce a compression scheme where certain subgraphs are
collapsed to reduce the number of edges at the cost of introducing an
additional vertex called a \emph{compressor vertex}.  The considered subgraph
is identical to a 3-stage Clos topology~\cite{clos1953study} and the
compression method removes the middle stage of vertices and instead introduces
the additional vertex connected to all vertices in the original
first and third Clos stage.
This work is extended~\cite{maccioni2016scalable} to cover other subgraphs that
can be replaced with sparser subgraphs.

\vspace{-0.25em}
\subsection{Compressing Associated Data Structures}

Several works compress various data structures related to the graph data
and used to, for instance, speed up some queries or for indexing purposes.
Ferragina et al.~\cite{ferragina2015compressed} compress indexing schemes for
large graphs; they target datasets that have vertices labeled with variable-length strings.
Jin et al.~\cite{jin2008efficiently} compress
transitive closures with spanning trees.

\vspace{-0.25em}
\subsection{Others}

Gbase~\cite{kang2011gbase, kang2012gbase} is a graph management system that
takes
as input a single big file with a list of edges and partitions it into several
homogeneous blocks.  Second, vertices are reshuffled, i.e., they are placed in
the blocks where the majority of their neighbors reside. Thanks to it, the
resulting blocks are either sparse or dense.  Next, Gbase compresses all
non-empty blocks through standard compression such as gzip. Finally, the
compressed blocks (and some meta information (e.g., the block row id) are
stored in the graph databases.

\begin{table*}
%\vspace{-1.5em}
\centering
\scriptsize
%\ssmall
\sf
\setlength{\tabcolsep}{2.5pt}
\begin{tabular}{lllllll}
\toprule
\textbf{Reference}$^{*}$ & \textbf{Size [bits]} & \makecell[c]{\textbf{Labels}\\$^{**}$} & \makecell[c]{\textbf{Edges}\\$^{***}$} & \makecell[c]{\textbf{Fast access /}\\ \textbf{encoding}$^{****}$} & \makecell[c]{\textbf{Targeted graph family}\\(see~\cref{sec:graph_families})} & \makecell[c]{\textbf{Scheme type}\\(see~\cref{sec:succ_concepts})} \\
\midrule
\multicolumn{7}{c}{\textbf{Targeted graphs: planar and planar-like (encoding schemes without fast navigation)}} \\
\midrule
%
%Itai~\cite{itai1982representation} &  $\frac{3}{2} n \log n + \mathcal{O}(n)$ & yes & undir. & no / yes & Triangulation & Minimize $\rwh{T}_\mathcal{A}(\mathcal{P},S)$ \\
Itai~\cite{itai1982representation} &  $\frac{3}{2} n \log n + \mathcal{O}(n)$ & yes & undir. & no / yes & Triangulation & Space/work-optimal \\
Turan~\cite{turan1984succinct} &  $\leq 12n$ & no & undir. & no / yes & Simple & \textasciitilde{}succinct \\
Turan~\cite{turan1984succinct} &  $n \lceil \log n \rceil + 12n$ & yes & undir. & no / yes & Simple & \textasciitilde{}succinct \\
Keeler~\cite{keeler1995short} & $n \log n + m \log 12 + o(n)$ & yes & undir. & no / yes & General & space-efficient \\
Keeler~\cite{keeler1995short} & $3m + \mathcal{O}(1)$ & no & undir. & no / yes & Map, stick-free, loop-free & space-efficient \\
Keeler~\cite{keeler1995short} & $m \log 12 + \mathcal{O}(1)$ & no & undir. & no / yes & General & space-efficient \\
Keeler~\cite{keeler1995short} & $m \log 12 + \mathcal{O}(1)$ & no & undir. & no / yes & Map, stick-free & space-efficient \\
Keeler~\cite{keeler1995short} & $m \log 12 + \mathcal{O}(1)$ & no & undir. & no / yes & Map, stick-free & space-efficient \\
Keeler~\cite{keeler1995short} & $(3 + \log 3)m/3 + \mathcal{O}(1)$ & no & undir. & no / yes & Triangulation & space-efficient \\
%
% & Itai~\cite{itai1982representation} &  $\frac{3}{2} n \log n + \mathcal{O}(n)$ & yes & undir. & no & yes & Planar triangulation \\
% & Tamassia~\cite{tamassia1988dynamic} & $\mathcal{O}(n)$ & ? & undir, $\leftarrow$ & yes & yes & Planar embedding \\
%
%
Chuang~\cite{chuang1998compact} &  $(n+m)\log3 + 1$ & no & undir. & no / yes & Simple, 3-connected & compact \\
Chuang~\cite{chuang1998compact} &  $(\min\{n,f\}+m)\log3 + 2$ & no & undir. & no / yes & Simple, 3-connected & compact \\
Chuang~\cite{chuang1998compact} &  $(n+m)\log3 + n + 1$ & no & undir. & no / yes & Simple, 3-connected, loops & compact \\
Chuang~\cite{chuang1998compact} &  $(\min\{n,f\}+m)\log3 + n + 2$ & no & undir. & no / yes & Simple, 3-connected, loops & compact \\
King~\cite{king1999guaranteed} &  $\frac{11}{3}n$ & yes & unsp. & no / yes & Planar, triangle, loop-free & \textasciitilde{}compact \\
He~\cite{he1999linear} & $4n-9 = \frac{4}{3}m - 1$ & no & undir. & no / yes & Plane triangulation & \textasciitilde{}succinct\\
He~\cite{he1999linear} & $(\frac{5}{2} + 2 \log 3) \min\{n,f\} - 7$ & no & undir. & no / yes & Plane 3-connected & \textasciitilde{}succinct \\
He~\cite{he2000fast} & $\beta(n) + o(\beta(n))$ & yes & both & no / yes & Plane triangulation & \textasciitilde{}succinct \\
He~\cite{he2000fast} & $\beta(n) + o(\beta(n))$ & yes & both & no / yes & Plane or planar & \textasciitilde{}succinct  \\
\makecell[l]{Poulalhon\\
\cite{poulalhon2003optimal, poulalhon2006optimal}} & $\log \binom{4n}{n} + o(n) \sim n \log \frac{256}{27} + o(n)$ & unsp. & unsp. & no / no & Triangulation & succinct \\
\makecell[l]{Fusy~\cite{fusy2005dissections}} & $\frac{1}{2} \log \left|\mathcal{P}(n)\right|$ & yes & unsp. & no / no & 3-connected, planar & succinct \\
\makecell[l]{Aleardi~\cite{aleardi2009schnyder}} & $4n + \mathcal{O}(g \log n)$ & yes & unsp. & no / yes & triangulation, genus $g$ & succinct \\
\makecell[l]{Aleardi~\cite{aleardi2010optimal}} & $\log \left| \mathcal{T}_{n',k}^{(\tau)} \right|$ & unsp. & unsp. & no / yes & \makecell[l]{triangulation,\\$n'$ inner vertices,\\$k$ boundary vertices} & succinct \\
\makecell[l]{Aleardi~\cite{aleardi2010optimal}} & $\log \left| \mathcal{Q}_{n',k}^{(\tau)} \right|$ & unsp. & unsp. & no / yes & \makecell[l]{bipartite quadrangulation,\\$n'$ inner vertices,\\$2k$ boundary vertices} & succinct \\
Despre~\cite{despre2017encoding} & $\log \binom{4n-2}{n-1} + o(n) \sim n \log \frac{256}{27} + o(n)$ & unsp. & unsp. & no / yes & toroidal triangulation & succinct \\
%
% \midrule
%
\bottomrule
\end{tabular}
\caption{(\cref{sec:succ_schemes}) Compact and succinct graph representations
for graphs that are \textbf{planar} or \textbf{planar-like} (maps, plane
graphs, etc.). $^{*}$ To save space, we only show the first name. $^{**}$
``yes'', ``no'', ``edges'' indicate that a graph has vertex labels, has no
labels at all, has edge labels; ``unsp.'' means labeling is not mentioned.
$^{***}$ ``undir.'', ``dir.'', ``both'' indicate that a scheme targets
undirected graphs, directed graphs, or both; ``unsp.'' means it is
unspecified.  $^{****}$ ``Fast'' indicates that a given scheme attempts to
reduce the time complexity of a certain query (queries) or the time to create
(i.e., encode) or decode a given representation from the input graph
representation (an AL or an AM); ``yes$\uparrow$'' indicates that the scheme
in the given row offers more efficient operations on the graph than the
corresponding scheme in the row \emph{below}, possibly at the cost of more
storage.  Note that in all the schemes with no or with unspecified labels,
one can attach information to vertices and edges using a simple array; the
cost of this operation is precisely the cost of the additional data.
Note that the sublinear component in time complexities in compact and succinct
schemes based on hierarchical decomposition (cf.~\cref{sec:succ_main_tool})
equals $o(X) = O\left(X \frac{\log \log X}{\log X} \right)$.
}
\label{tab:succ_planar_1}
\end{table*}

\begin{table*}
%\vspace{-1.5em}
\centering
\scriptsize
%\ssmall
\sf
\setlength{\tabcolsep}{2.5pt}
\begin{tabular}{lllllll}
\toprule
\textbf{Reference}$^{*}$ & \textbf{Size [bits]} & \makecell[c]{\textbf{Labels}\\$^{**}$} & \makecell[c]{\textbf{Edges}\\$^{***}$} & \makecell[c]{\textbf{Fast access /}\\ \textbf{encoding}$^{****}$} & \makecell[c]{\textbf{Targeted graph family}\\(see~\cref{sec:graph_families})} & \makecell[c]{\textbf{Scheme type}\\(see~\cref{sec:succ_concepts})} \\
\midrule
\multicolumn{7}{c}{\textbf{Targeted graphs: planar and planar-like (encoding schemes with fast navigation)}} \\
\midrule
Tamassia~\cite{tamassia1988dynamic} & $\mathcal{O}(n)$ & unsp. & both & yes / yes & Planar embedding & compact \\
\makecell[l]{Munro\\\cite{munro2001succinct,munro1997succinct}} &  $8n + 2m + o(n+m)$ & yes & undir. & yes / no & General & succinct \\
Chuang~\cite{chuang1998compact} &  $2m + \left(5 + \frac{1}{k}\right)n + o(m+n)$ & no & undir. & yes$\uparrow$ / yes & General, loop-free, $k > 0$ & compact \\
Chuang~\cite{chuang1998compact} &  $2m + \frac{14}{3} n + o(m+n)$ & no & undir. & yes / yes & General, loop-free & compact \\
Chuang~\cite{chuang1998compact} &  $\frac{5}{3} m + \left(5 + \frac{1}{k}\right)n + o(n)$ & no & undir. & yes$\uparrow$ / yes & Simple, $k > 0$ & compact \\
Chuang~\cite{chuang1998compact} &  $\frac{4}{3} m + 5n + o(n)$ & no & undir. & yes / yes & Simple & compact \\
Chuang~\cite{chuang1998compact} &  $2m + 3n + o(m+n)$ & no & undir. & yes / yes & General, 3-connected, loop-free & compact \\
Chuang~\cite{chuang1998compact} &  $2m + 2n + o(n)$ & no & undir. & yes / yes & Simple, 3-connected & compact \\
Chuang~\cite{chuang1998compact} &  $2m + 2n + o(m+n)$ & no & undir. & yes / yes & General, triangulated, loop-free & compact \\
Chuang~\cite{chuang1998compact} &  $2m + n + o(n)$ & no & undir. & yes / yes & Simple, triangulated & compact \\
Chuang~\cite{chuang1998compact} &  $2m + \left(6 + \frac{1}{k}\right)n + o(m+n)$ & no & undir. & yes$\uparrow$ / yes & General, $k > 0$ & compact \\
Chuang~\cite{chuang1998compact} &  $2m + \frac{17}{3} n + o(m+n)$ & no & undir. & yes / yes & General & compact \\
Chuang~\cite{chuang1998compact} &  $\frac{5}{3} m + \left(6 + \frac{1}{k}\right)n + o(n)$ & no & undir. & yes$\uparrow$ / yes & Simple, loops, $k > 0$ & compact \\
Chuang~\cite{chuang1998compact} &  $\frac{4}{3} m + 6n + o(n)$ & no & undir. & yes / yes & Simple, loops & compact \\
Chuang~\cite{chuang1998compact} &  $2m + 4n + o(m+n)$ & no & undir. & yes / yes & General, 3-connected & compact \\
Chuang~\cite{chuang1998compact} &  $2m + 3n + o(n)$ & no & undir. & yes / yes & Simple, 3-connected, loops & compact \\
Chuang~\cite{chuang1998compact} &  $2m + 3n + o(m+n)$ & no & undir. & yes / yes & General, triangulated & compact \\
Chuang~\cite{chuang1998compact} &  $2m + 2n + o(n)$ & no & undir. & yes / yes & Simple, triangulated, loops & compact \\
Chiang~\cite{Chiang:2001:OST:365411.365518} &  $2m + 3n + o(m+n)$ & no & undir. & yes / yes & General, loop-free & space-efficient \\
Chiang~\cite{Chiang:2001:OST:365411.365518} &  $2m + 2n + o(n)$ & no & undir. & yes / yes & Simple, loop-free & space-efficient \\
%
%
%
% Aleardi~\cite{aleardi2005succinct, aleardi2006compact} &  $2.175t + O\left(t \frac{\log \log t }{ \log t}\right)$ & unsp. & unsp. & yes / yes & Triangulation & succinct \\ 
% Aleardi~\cite{aleardi2005succinct, aleardi2006compact} &  \makecell[l]{$2.175t + 36(g-1)\log t$\\ $+ O\left(t \frac{\log \log t }{ \log t} + g \log \log t \right)$} & unsp. & unsp. & yes / yes & \makecell[l]{Triangulation of\\a surface with genus $g$} & succinct \\ 
%
Aleardi~\cite{aleardi2005succinct, aleardi2006compact} &  $2.175t + o\left(t\right)$ & unsp. & unsp. & yes / yes & Triangulation & succinct \\ 
Aleardi~\cite{aleardi2005succinct, aleardi2006compact} &  \makecell[l]{$2.175t + 36(g-1)\log t$\\ $+ o(t) + \mathcal{O}\left(g \log \log t \right)$} & unsp. & unsp. & yes / yes & \makecell[l]{Triangulation of\\a surface with genus $g$} & succinct \\ 
Aleardi~\cite{aleardi2005published} & $2.17m + \mathcal{O}(g \log m) + o(m)$ & unsp. & unsp. & yes / no & Dynamic triangulation & succinct \\
Barbay~\cite{labeled, barbay2012succinct} &  $2m \log 6 + o(m)$ & no & unsp. & yes / no & Planar triangulation & succinct \\
Barbay~\cite{labeled, barbay2012succinct} &  $t \log \sigma + t \cdot o(\log \sigma)$ & no & unsp. & yes / no & Triangulation & succinct \\
Barbay~\cite{labeled, barbay2012succinct} &  $t \log \sigma + t \cdot o(\log \sigma)$ & yes & unsp. & yes / no & Triangulation & succinct \\
Barbay~\cite{labeled, barbay2012succinct} &  $n + t(\log \sigma + o(\log \sigma))$ & edges & unsp. & yes / no & Outerplanar & succinct \\
Barbay~\cite{labeled, barbay2012succinct} &  $4n + t(\log \sigma + o(\log \sigma))$ & edges & unsp. & yes / no & General & succinct \\
Aleardi~\cite{aleardi2008succinct} &  $2m + o(n)$ & unsp. & unsp. & yes / yes & 3-connected & succinct \\
Aleardi~\cite{aleardi2008succinct} &  $3.24n + o(n)$ & unsp. & unsp. & yes / yes & Triangulated & succinct \\
Blelloch~\cite{blelloch2010succinct} &  $\mathcal{H}_p(n) + o(n)$ & no & unsp. & yes / yes & Map & succinct \\
Yamanaka~\cite{Yamanaka:2008:CEP:1787651.1787667} & $6n + o(n)$ & yes & unsp. & yes / no & Plane triangulation & compact \\
\bottomrule
\end{tabular}
\caption{(\cref{sec:succ_schemes}) Compact and succinct graph representations
for graphs that are \textbf{planar} or \textbf{planar-like} (maps, plane
graphs, etc.). $^{*}$ To save space, we only show the first name. $^{**}$
``yes'', ``no'', ``edges'' indicate that a graph has vertex labels, has no
labels at all, has edge labels; ``unsp.'' means labeling is not mentioned.
$^{***}$ ``undir.'', ``dir.'', ``both'' indicate that a scheme targets
undirected graphs, directed graphs, or both; ``unsp.'' means it is
unspecified.  $^{****}$ ``Fast'' indicates that a given scheme attempts to
reduce the time complexity of a certain query (queries) or the time to create
(i.e., encode) or decode a given representation from the input graph
representation (an AL or an AM); ``yes$\uparrow$'' indicates that the scheme
in the given row offers more efficient operations on the graph than the
corresponding scheme in the row \emph{below}, possibly at the cost of more
storage.  Note that in all the schemes with no or with unspecified labels,
one can attach information to vertices and edges using a simple array; the
cost of this operation is precisely the cost of the additional data.
Note that the sublinear component in time complexities in compact and succinct
schemes based on hierarchical decomposition (cf.~\cref{sec:succ_main_tool})
equals $o(X) = O\left(X \frac{\log \log X}{\log X} \right)$.
}
\label{tab:succ_planar_2}
\end{table*}

\begin{table*}
\centering
\scriptsize
\sf
\setlength{\tabcolsep}{2.5pt}
\begin{tabular}{lllllll}
\toprule
\textbf{Reference}$^{*}$ & \textbf{Size [bits]} & \makecell[c]{\textbf{Labels}\\$^{**}$} & \makecell[c]{\textbf{Edges}\\$^{***}$} & \makecell[c]{\textbf{Fast access /}\\ \textbf{encoding}$^{****}$} & \makecell[c]{\textbf{Targeted graph}\\(family; see~\cref{sec:graph_families})} & \makecell[c]{\textbf{Scheme type}\\(see~\cref{sec:succ_concepts})} \\
\midrule
\multicolumn{7}{c}{\textbf{Targeted graphs: ``middle-ground'' ($k$-page, separable) that are more general than the planar ones}} \\
\midrule
\makecell[l]{Jacobson\\\cite{Jacobson:1988:SSD:915547, jacobson1989space}} &  $\mathcal{O}(kn)$ & no & unsp. & yes / yes & $k$-page & succinct \\
Cohen~\cite{Cohen:1993:RWO:167088.167152} & $\mathcal{O}(n)$ [``space''] & yes & unsp. & yes / yes & $k$-connected & \textasciitilde{}compact \\
\makecell[l]{Munro\\\cite{munro2001succinct,munro1997succinct}} &  $2kn + 2m + o(kn + m)$ & no & yes & & $k$-page & succinct \\
Deo~\cite{deo1998structural} & $\mathcal{O}(n+g)$ & yes & undir. & no / yes & bounded genus ($\le g$) & \textasciitilde{}compact \\
Deo~\cite{deo1998structural} & $\mathcal{O}(n)$ & yes & undir. & no / yes & \makecell[l]{bounded arboricity,\\ separable} & \textasciitilde{}compact \\
Lu~\cite{lu2002linear, lu2014linear} & $\le \beta(n) + o(\beta(n))$ & yes & undir. & no / yes & \makecell[l]{genus $g = \left( \frac{n}{\log^2 n} \right)$, \\ others (see~\cite{lu2002linear, lu2014linear})} & \textasciitilde{}compact \\
Blandford~\cite{Blandford:2003:CRS:644108.644219} & $\mathcal{O}(n)$ & no & both & yes / no & separable & compact \\
Barbay~\cite{labeled, barbay2012succinct} & $n + 2m \log k + o(m \log k)$ & no & unsp. & yes / no& $k$-page & succinct \\
Barbay~\cite{labeled, barbay2012succinct} & $n + (2+\epsilon)m \log k + o(m \log k)$ & no & unsp. & yes / no & $k$-page & succinct \\
Barbay~\cite{labeled, barbay2012succinct} & $kn + t(\log \sigma + o(\log \sigma))$ & edges & unsp. & yes / no & $k$-page & succinct \\
Barbay~\cite{labeled, barbay2012succinct} & \makecell[l]{$n + (2m + \epsilon) \log k + o(m \log k)$\\ $+ m(\log \sigma + o(\log \sigma))$} & edges & unsp. & yes / no & $k$-page & succinct \\
Gavoille~\cite{gavoille2008compact} & $2m \log k + 4m$ & no & undir. & no / no & $k$-page, $k \le \frac{1}{2} kn / \log k$ &  \textasciitilde{}compact \\
Gavoille~\cite{gavoille2008compact} & $2m \log k + 4m + o(m \log k)$ & no & undir. & yes$\uparrow$ / no & $k$-page, $k \le \frac{1}{2} kn / \log k$ & \textasciitilde{}compact \\
Gavoille~\cite{gavoille2008compact} & $2m \log k + 4m + o(m)$ & no & undir. & yes / no & $k$-page, $k \le \frac{1}{2} kn / \log k$ & \textasciitilde{}compact \\
Blelloch~\cite{blelloch2010succinct} &  $\mathcal{H}(n) + o(n)$ & no & yes & yes / no & separable & succinct \\
\midrule
\multicolumn{7}{c}{\textbf{Targeted graphs: arbitrary (no or little assumptions on the structure)}} \\
\midrule
Turan~\cite{turan1984succinct} &  $\binom{n}{2} - \frac{1}{8} n \log n + \mathcal{O}(n)$ & no & unsp. & no / no  & - & \textasciitilde{}succinct \\
Naor~\cite{naor1990succinct} & $\binom{n}{2} - n \log n + \mathcal{O}(n)$ & no & unsp. & yes / yes & \makecell[l]{arbitrary adjacency\\\textbf{matrix}} & \makecell[l]{Approach the\\AM bound $\binom{n}{2}$ \\ $- n \log n + \mathcal{O}(n)$} \\
Raman~\cite{raman2007succinct} & $\left\lceil \log \binom{n^2}{m} \right\rceil + o(m)$ & no & dir. & yes / no & - & succinct \\
Farzan~\cite{farzan2008succinct} &  $\log \binom{n^2}{m} + o\left(\log \binom{n^2}{m}\right)$ & yes & dir. & yes / no & $m > \frac{n^2}{\log^{1/3} n}$ & succinct \\
Farzan~\cite{farzan2008succinct} &  $(1+\epsilon) \log \binom{n^2}{m}$ & yes & dir. & yes / no & $\frac{n^2}{\log^{1/3} n} \ge m > \frac{n}{2}$ & succinct \\
Farzan~\cite{farzan2008succinct} &  $\log \binom{n^2}{m} + \epsilon m \log m$ & yes & dir. & yes / no & $\epsilon > 0$, $m \le \frac{n}{2}$ & succinct \\
Farzan~\cite{farzan2008succinct} &  $\log \binom{n^2/2}{m}$ & yes & undir. & yes / no & $ \frac{n^2}{4} > m > \frac{n^2}{\log^{1/3} n}$ & succinct \\
Farzan~\cite{farzan2008succinct} &  $(1+\epsilon) \log \binom{n^2/2}{m}$ & yes & undir. & yes / no & $\frac{n^2}{\log^{1/3} n} \ge m > \frac{n}{2}$ & succinct \\
Farzan~\cite{farzan2008succinct} &  $\log \binom{n^2/2}{m} + \epsilon m \log m$ & yes & undir. & yes / no & $\epsilon > 0$, $m \le \frac{n}{2}$ & succinct \\
Fischer~\cite{fischer2016glouds} &  \makecell[l]{$(2n+m)\log 3 + h \log n + k \log h $\\$+o(m + k\log h) + \mathcal{O}(\log \log n)$} & unsp. & dir. & yes / no & \makecell[l]{$k = \mathcal{O}(n), k=m-n+1$ \\ $h = |K| \le k$\\$K = \{v \in V: |N_{in,v}| > 1\}$} & succinct \\
\bottomrule
\end{tabular}
\caption{(\cref{sec:succ_schemes}) Compact and succinct graph
representations for \textbf{``middle-ground''} and \textbf{arbitrary} graphs.
$^{*}$ To save space, we only show the first
name.  $^{**}$ ``yes'', ``no'', or ``edges'' indicate that a graph has vertex
labels, has no labels at all, or has edge labels; ``unsp.'' means labeling is
not mentioned.  $^{***}$ ``undir.'', ``dir.'', and ``both'' indicate that a
scheme targets undirected graphs, directed graphs, or both; ``unsp.'' means it
is unspecified.  $^{****}$ ``Fast'' indicates that a given scheme attempts to
reduce the time complexity of a certain query (or queries) or the time to
create (i.e., encode) or decode a given representation from the input graph
representation (an AL or an AM); ``yes$\uparrow$'' indicates that the scheme 
in the given row offers more efficient operations on the graph than the
corresponding scheme in the \emph{below}, possibly at the cost of more storage.
Note that in all the schemes with no or with unspecified labels, one can
attach information to vertices and edges using a simple array; the cost
of this operation is precisely the cost of the additional data.
}  
\vspace{-2em}
\label{tab:succ_middle}
\end{table*}

\section{Approaching Storage Lower Bounds}
\label{sec:bounds}

The core idea behind these schemes is to encode a given graph so that the
representation explicitly approaches the storage lower bound. Unless stated
otherwise, all schemes operate on static graphs. 
Tables~\ref{tab:succ_planar_1}, \ref{tab:succ_planar_2}, and~\ref{tab:succ_middle} feature the considered
representations together with storage complexities.

% \macb{Scope}
% %
% Here, we omit a detailed description of works dedicated to triangulations,
% meshes, and any types of planar graphs, and only provide complexity results for
% completeness. This reason is that these schemes fall partially into the
% category of meshes covered in existing surveys~\cite{d,s,d}. 
%
% In addition, the need for compressing such graphs is of lower importance as
% real-world datasets belonging to such families are usually of low size (e.g.,
% the largest USA roadmap has only $\approx$30M edges~\cite{where}).

\subsection{Related Concepts}
\label{sec:succ_concepts}

We first explain all the related notions: \emph{succinct} or \emph{compact}
graph representations. We define them in~\cref{sec:succ_defs}.
% 
% using
% definitions (common in the related literature) based on optimum bit counts.
% %
% We then discuss an alternative way of defining succinct structures via entropy
% and the relationship between these two terms in~\cref{sec:succ_entropy}.
%
Next, these terms are sometimes used imprecisely; we clarify such issues and
provide a taxonomy of these terms used in this survey
in~\cref{sec:succ_final_taxo}. We also illustrate the main techniques used in
achieving succinctness in the majority of graph representations and
others in~\cref{sec:succ_main_tool}.

%
% =======
% \emph{succinct}, \emph{compact}, or \emph{implicit}. These are precise terms
% that we clarify in~\cref{sec:succ_defs} using
% definitions based on \emph{optimum bit counts}.
% %
% We then discuss an alternative way of defining succinct structures via entropy~\cref{sec:succ_entropy}.
% %
% Third, all these terms are often used imprecisely;
% we clarify these issues
% %
% %in~\cref{sec:succ_clarify}--\cref{sec:impl_clarify}
% %
% >>>>>>> 20635b3fc6289de449796eaa530cb2a42e1e688d
%

% \subsubsection{Succinctness, Compression, Condensing}
% \label{sec:succinctness_background}

% Succinctness and compression are two complementary mechanisms to reduce the
% size of $G$.

\subsubsection{Succinctness and Compactness: Definitions}
\label{sec:succ_defs}

% We first give the definitions of succinctness, compactness, and implicitness,
% based on counting  
% 
% We first summarize related concepts.
%
Assume $\mathcal{N}$ is the optimal bit count to store some data. 
A representation of this data can be \emph{compact} 
or \emph{succinct}.
In the former case, it uses
$\mathcal{O}(\mathcal{N})$ bits; in the latter case, it uses $\mathcal{N} +
o(\mathcal{N})$ bits. These definitions are used in various modern works on
succinct data structures~\cite{aleardi2008succinct, succ-category,
munro2004succinct}.
Many of these representations simultaneously support \emph{a reasonable set of
queries fast} (e.g., in $\mathcal{O}(1)$ or $\mathcal{O}(\log n)$
time)~\cite{Blandford:2003:CRS:644108.644219}, others only consider reducing
space complexity. Finally, some designs accelerate the process of generating
(encoding) a given data representation.

Specifically, when considering an arbitrary
(undirected) graph with $n$ vertices and $m$ edges, the number of such graphs
is $\mathcal{N} = \binom{\binom{n}{2}}{m}$, the storage lower bound
is $\left \lceil \log \mathcal{N} \right \rceil$, and thus a succinct and
compact representation respectively take $\left \lceil \log \mathcal{N} \right
\rceil + o(\log \mathcal{N})$ and $\mathcal{O}(\log \mathcal{N})$ bits.

% \subsubsection{Succinctness: Definition Based on Entropy}
% \label{sec:succ_entropy}

One can equivalently define succinctness via entropy. For example, Aleardi et
al.~\cite{aleardi2005succinct, aleardi2008succinct} state that a data structure
is succinct if its asymptotic size \emph{``matches the entropy of the class [of
represented structures] and compact if it matches it up to a constant factor''}. 
In the language of the terms used above, the class of represented structures
are arbitrary graphs.

% we consider the number of objects of size n in the class. If this number has
% an exponential growth of order 2 a n when n goes to infinity, the entropy1 of
% the class is defined to be a n. A representation is then compact if it uses
% O(n) bits and succinct if it uses a n + o(n) bits. Observe that a correct
% representation cannot use less than a n bits, for it must be possible to
% distinguish all objects.

% we consider the number of objects of size n in the class. If this number has
% an exponential growth of order 2 a n when n goes to infinity, the entropy1 of
% the class is defined to be a n. A representation is then compact if it uses
% O(n) bits and succinct if it uses a n + o(n) bits. Observe that a correct
% representation cannot use less than a n bits, for it must be possible to
% distinguish all objects.

%\subsubsection{Succinctness and Compactness: Clarifications}
%\label{sec:succ_clarify}

% \subsubsection{Implicitness: Clarifications}
% \label{sec:impl_clarify}

\subsubsection{Succinctness and Compactness: Taxonomy of Concepts}
\label{sec:succ_final_taxo}

We now clarify certain issues related to succinctness and compactness. Recent
works associate these terms with the definitions in~\cref{sec:succ_defs}. 
Now, there exist various works that use different, although related,
definitions of succinctness and compactness. For example, Galperin and
Wigderson state that a succinct graph representation takes $o(n)$
space~\cite{galperin1983succinct}.  We now describe these aspects and introduce
notions that are used in the remainder of this section and in
Table~\ref{tab:succ_planar_1}, \ref{tab:succ_planar_2}, and~\ref{tab:succ_middle}. Specifically, we use the
following terms:

\noindent
\macb{succinct}
This term indicates that a representation uses the definition of succinctness
from~\cref{sec:succ_defs} or the definition of succinctness based on
entropy~\cite{aleardi2005succinct, aleardi2008succinct} provided in the last
paragraph of~\cref{sec:succ_defs}.

\noindent
\macb{\textasciitilde{}succinct}
An encoding of a graph $G$ from a class of graphs $\mathcal{G}$ is described
with this term if it is described as succinct but the definition does not match
the one from~\cref{sec:succ_defs} or the one based on entropy and stated in
past work~\cite{aleardi2005succinct, aleardi2008succinct}. Example such
definitions are
``the length of the encoding of $G$'s representation is not too large compared
to $\log|\mathcal{G}|$''~\cite{turan1984succinct},
``the length of this encoding is not much larger than its information-theoretic
tight bound, i.e., the shortest length over all possible coding
schemes''~\cite{he1999linear}.

\noindent
\macb{compact}
This term indicates that a representation uses the definition of
compactness from~\cref{sec:succ_defs}.

\noindent
\macb{\textasciitilde{}compact}
This term describes a representation that does not explicitly use the
definition of compactness from~\cref{sec:succ_defs} but directly compares to
and discusses such representations~\cite{chuang1998compact}.

\noindent
\macb{space-efficient}
We use this term when a given representation claims relationships to succinctness or
compactness, and/or extensively discusses succinct or compact representations,
but when it is not itself based on any precise optimality conditions. For
example, Chiang et al.~\cite{Chiang:2001:OST:365411.365518} states that a
``space-efficient'' representation (1) minimizes the length of a given
encoding, (2) minimizes the time required to compute and decode the encoding,
and (3) supports queries on this encoding efficiently.

\subsubsection{Main Techniques for Achieving Succinctness}
\label{sec:succ_main_tool}

We now describe generic techniques for obtaining succinctness and compactness
that are often used in various representations.

\macb{Hierarchical Decomposition}
Assume that the size of a considered class of objects is
$\mathcal{N}$. For example, for a class of arbitrary graphs we have
$\mathcal{N} = \binom{\binom{n}{2}}{m}$.
The high-level key idea is to divide the object to be encoded (e.g., an
arbitrary graph) into \emph{small} parts, group these parts in an auxiliary
table, and represent them with the indices into this table. This table should
contain \emph{all} possible parts so that \emph{any} object from a given class
could be constructed from them. These parts are again divided into yet smaller
(\emph{tiny}) parts, stored similarly in yet other auxiliary tables.  Now, the
size of both small and tiny parts is selected in such a way that the sum of the
sizes of all the indices and all the auxiliary tables is
$\mathcal{O}(\mathcal{N})$ (for compactness) and $\mathcal{N} + o(\mathcal{N})$
(for succinctness) bits.
The central observation that enables these bounds is that the representation
consisting of small and tiny parts can be \emph{hierarchical}: tiny parts only
need pointers that point to other tiny parts within a \emph{single} small part
because small parts use other pointers that link them to other small parts.

More formally, an object to be encoded (e.g., a graph) is first divided into
\emph{tiny} parts of size $\mathcal{O}(\log \mathcal{N})$. Here, \emph{tiny}
means that the catalog of all these parts (i.e., an auxiliary table explicitly
storing these parts) must take $o(\mathcal{N})$; in most cases it is
$\mathcal{O}\left(\frac{\mathcal{N}}{\log \mathcal{N}} \right)$. 
The index in this catalog of a given tiny part and the sum of the sizes of all such
indices represent together this part.

Second, one must encode how these tiny parts together form the initial object.
Now, the number of these parts is $\mathcal{O}\left(\frac{\mathcal{N}}{\log
\mathcal{N}} \right)$, the number of connections between them is
$\mathcal{O}(\mathcal{N})$, and a pointer to any such part takes
$\mathcal{O}(\log \mathcal{N})$ bits. Thus, a classical representation of the
connections between tiny parts is $\mathcal{O}(\mathcal{N})$, \emph{giving a
compact representation}.

To achieve succinctness, $\log \mathcal{N}$ tiny parts are combined into
\emph{small} parts, each of which uses $\mathcal{O}(\log^2 \mathcal{N})$ space.
Thus, pointers between small parts can have size of $\mathcal{O}(\log \mathcal{N})$
while tiny parts can use pointers of size $\mathcal{O}(\log \log
\mathcal{N})$ because they now need to point to each other only within one
small part. Now, the total count of small and tiny parts is
$\mathcal{O}\left(\frac{\mathcal{N}}{\log^2 \mathcal{N}} \right)$ and
$\mathcal{O}\left(\frac{\mathcal{N}}{\log \mathcal{N}} \right)$, respectively.
This gives the total size of
$\mathcal{O}\left(\frac{\mathcal{N}}{\log\mathcal{N}}\right)$ bits and
$\mathcal{O}\left(\frac{\mathcal{N} \log \log \mathcal{N}}{\log
\mathcal{N}}\right)$ bits, respectively.

\macb{Parentheses Encoding}
Another general succinct or compact encoding uses parentheses. To explain the idea
intuitively, consider a tree and a Depth-First Search traversal of this tree.
One can represent this tree with a string consisting of two parentheses, ``(''
and ``)''.  Namely, during the traversal, when a vertex is visited for the
first time, one appends an opening parenthesis ``('' to the string. When a
vertex is visited for the second and the last time (while moving backwards in
the tree structure), the other parenthesis ``)'' is added.  Thus, a tree with
$n$ vertices uses $2n$ bits (one bit per one parenthesis type).
Now, a graph could be represented in a similar way. For example, if 
a graph is decomposed into a set of spanning trees, each tree could be represented
with such a string of parentheses that are in practice encoded with ``0''s and
``1''s.

\subsection{Succinct and Compact Schemes}
\label{sec:succ_schemes}

We next describe several concrete succinct and compact representations. We
summarize all the considered schemes in Table~\ref{tab:succ_planar_1}, \ref{tab:succ_planar_2}, 
and~\ref{tab:succ_middle}.  The former presents planar graphs, subclasses of
planar graphs, and planar-related ones such as maps. The latter summarizes
``middle-ground'' graphs and graphs of arbitrary structure; ``middle-ground''
are graphs that are more generic than planar ones (edges can cross outside
their adjacent vertices), but do have some strong assumptions on the structure,
including bounded genus, bounded arboricity, bounded number of pages, or
separability. These classes are explained in~\cref{sec:graph_families}.
Finally, arbitrary graphs are graphs with any structure where the only
assumption can be related to the number of edges.
%
% In each table, we first give a reference to the associated paper (due to
% space constraints we only provide the first author).  Second, we privide the
% size of the associated representation [bits].  In the ``Labels'' column we
% state whether Other columns include: the presence of vertex or edge labels,

% \subsubsection{Planar-Related Graphs}
% \label{sec:succ_planar}

The majority of succinct and compact schemes listed in Tables~\ref{tab:succ_planar_1}, \ref{tab:succ_planar_2},
and~\ref{tab:succ_middle} use one of a few ``standard'' mechanisms for achieving 
compactness or succinctness, described in~\cref{sec:succ_main_tool}.
We now group these schemes basing on the associated mechanism.
% 
% and devote more
% detailed descriptions to selected papers.

\subsubsection{Schemes Based on Hierarchy}

There are numerous succinct and compact representations that use the generic
hierarchical way of obtaining the storage lower bounds.  They include planar
graphs~\cite{aleardi2005succinct, aleardi2008succinct, blelloch2010succinct},
``middle-ground'' graphs~\cite{blelloch2010succinct}, and arbitrary
graphs~\cite{Farzan:2013:SEA:2562345.2562632__}.  The key idea is as stated
in~\cref{sec:succ_main_tool}: all possible parts of any input graph (in a given
class) are indexed in a lookup table and pointers of specially engineered sizes
are used to ensure that any input graph can be constructed from the indexed
elements to provide the desired storage bounds.

\subsubsection{Schemes Based on Parentheses}

Many representations use the concept of parentheses for succinct or compact
encoding~\cite{jacobson1989space, munro1997succinct, gavoille2008compact, ferres2017fast}. In
some cases (usually planar graphs), they use one type of parentheses, but
several schemes propose to use multiple types of parentheses if the input graph
has a more complex structure.
Moreover, several other representations use parentheses combined with reordering the
vertices according to a special order called the \emph{canonical
order}~\cite{kant1996drawing, FraysseixPP90}. These are all planar
graphs~\cite{chuang1998compact, he1999linear, Chiang:2001:OST:365411.365518,
barbay2012succinct}.

\subsubsection{Schemes Based on Encoding Trees}

Some representations are based on decomposing the input graph into trees
constructed from a DFS graph traversal, and then encoding such trees using a 
selected scheme~\cite{keeler1995short, king1999guaranteed}.

\subsubsection{Others}

There are also other representations~\cite{itai1982representation, turan1984succinct,
viana2008quick, aleardi20062d, fusy2005dissections}.
For example,
Blandford et al.~\cite{Blandford:2003:CRS:644108.644219} relabel vertices based
on recursive partitioning of the input graph to achieve compactness.  Raman et
al.~\cite{raman2007succinct} use succinct indexable dictionaries as a basis for
ensuring succinct binary relations that can then be used to encode succinctly
an arbitrary graph. 
Various works discuss storage lower bounds of families of graphs such as
planar graphs~\cite{bonichon2006planar}.
Some encoding schemes
use Schnyder Woods, a closely related combinatorial structure~\cite{aleardi2009schnyder, Schnyder90}.
Elias-Fano code~\cite{vigna2013quasi, ottaviano2014partitioned} is a
``quasi-succinct'' encoding of sequences of integers that grow monotonically.
When traversing a sorted adjacency list, it enables proceeding (in almost
constant time) to the next neighbor that has the ID larger than a given value.
Thus, one can compute intersections of neighborhoods sublinearly.

\subsection{Other Storage Lower Bound Measures}
\label{sec:entropy}

Besides compactness and succinctness, there exist other concepts related to
storage lower bounds that could be used while developing and analyzing
compression schemes or storage-efficient representations. 
A detailed description of such concepts is outside the scope of this work.
However, we briefly mention them to make this survey complete and to provide
the associated links. One obvious related notion in discussing storage lower
bounds is graph entropy; it was covered in several
surveys~\cite{simonyi1995graph, simonyi2006entropies, mowshowitz2012entropy,
holzinger2013graph, dehmer2011history}. Another way to describe storage bounds
is Kolmogorov complexity of graphs that was covered in some
works~\cite{hearn2006applications, li2008p, ming1997introduction}.
These works most often focus on investigating the ``information content'' of a
given graph family, for example the notion of topological
entropy~\cite{rashevsky1955life, trucco1956note} is related to the probability
of a graph having a certain partitioning structure.
Chierichetti at el.~\cite{chierichetti2009models} discuss the information
content of web graphs and propose a graph model that reflects this content.

Some of these works specifically address
compression~\cite{choi2009compression, choi2010fast, choi2012compression,
guler2014study}.
Choi and Szpankowski~\cite{choi2009compression, choi2010fast,
choi2012compression} propose the ``Structural zip'' algorithm for compressing
unlabeled graphs; it compresses a given labeled $G$ into a codeword that can be
decoded into a graph isomorphic to $G$. 
The main idea behind the algorithm is as follows. First, a vertex $v_1$ is
selected and its neighbor count is stored explicitly. Then, the remaining $n-1$
vertices are partitioned into two sets: $v_1$'s neighbors and non-neighbors.
This continues recursively by selecting a vertex $v_2$ from $v_1$'s neighbors
and storing two numbers: the number of $v_2$ neighbors among each of these two
sets. Next, the remaining $n-2$ vertices are partitioned into four further
sets: the neighbors of both $v_1$ and $v_2$, the neighbors of $v_1$ that are
non-neighbors of $v_2$, the non-neighbors of $v_1$ that are $v_2$'s neighbors,
and the non-neighbors of both $v_1$ and $v_2$. This continues until all
vertices are processed.  During the algorithm execution, two types of encoded
neighbor counts are maintained and concatenated into one of the separate binary
sequences.  First, the neighbor counts that have more than one bit in length (i.e., for
subsets $|U| > 1$) are concatenated to form the first sequence.  Second, the
neighbor counts that have exactly one bit in length (i.e., for subsets $|U| = 1$) are
concatenated to form the second sequence.

Moreover, Luczak et al.~\cite{luczak2017structural} design asymptotically
optimal algorithms for compressing unlabeled and labeled graphs constructed
with the preferential attachment model.
Others analyze theoretical aspects of compressing clustered
graphs~\cite{abbe2016graph, asadi2017compressing} or establish a formal
relationship between the
storage lower bound of a given graph
\cite{litow2004graph}
and the polynomials with simple zeros.

\subsection{Discussions on Computational Complexity}

Some papers propose schemes for fast construction of succinct or compact
representations~\cite{ferres2017parallel} or discuss their computational
complexity (e.g., prove NP-hardness)~\cite{papadimitriou1986note,
galperin1983succinct}.

% Some papers discuss the computational complexity of deriving succinct representations,
% for example by showing NP-hardness of some schemes~\cite{papadimitriou1986note, galperin1983succinct}.

\section{Graph Minimum Arrangement for Storage Reductions}
\label{sec:relabeling}

Another line of works uses Integer Linear Programming (ILP) formulations to
compress graphs by reordering vertex labels such that the new labels can be
compressed more effectively.  For example, some schemes assign labels to
decrease differences between IDs of consecutive neighbors in each neighborhood;
these minimized differences are then encoded using some variable-length coding,
ultimately reducing the size of each such neighborhood and thus $G$'s size. 
We already discussed some schemes that aim at improving such reorderings; now
we focus on existing research that explicitly uses ILP formulations or improves
them.
This particular problem is called \emph{Minimum Linear Gap Arrangement}
(MLinGapA) because it consists in minimizing linear gaps between consecutive neighbors.
There are three other related problems: \emph{Minimum Logarithmic Gap
Arrangement} (MLogGapA), \emph{Minimum Linear Arrangement} (MLinA), and
\emph{Minimum Logarithmic Arrangement} (MLogA).
More generally, this family of problems is called \emph{Minimum Arrangement}
problems and are a part of a domain called \emph{Graph Layout}
problems~\cite{diaz2002survey}.

\subsection{Definitions of Minimimum Arrangement Problems}
\label{sec:ilp_defs}

Formally, a \emph{layout} of an undirected graph $G$ is a bijective function
$\phi : V \to [n] = \{1, ... , n\}$~\cite{petit2013addenda} that reassigns labels of vertices
so that a certain function is minimized. 
%
% Now, given a
% layout $\phi$ of $G$ and an integer $i$, we define two sets:
% 
% \begin{gather}
% %
% L(i, \phi, G) = {u \in V | \phi(u) \le i} \\
% R(i, \phi, G) = {u \in V | \phi(u) > i} 
% %
% \end{gather}
% 
% % Then, the \emph{edge cut} at position $i$ of $\phi$ is defined as
% % 
% % \begin{gather}
% % %
% % \theta(i, \phi, G) = |\{uv \in E | u \in L(i, \phi ,G) \land v \in R(i, \phi, G)\}|
% % %
% % \end{gather}
% 
% Next, for a layout $\phi$ of $G$ and an edge $uv \in E$, the \emph{length}
% of $uv$ on $\phi$ is
% 
% \begin{gather}
% %
% \lambda(uv, \phi ,G) = |\phi(u) - \phi(v)|
% %
% \end{gather}
% 
% Now, a \emph{layout cost} is a function $F$ that associates an integer number
% $F(\phi, G)$ with each layout $\phi$ of a graph $G$. The optimization layout
% problem associated with $F$ aims at finding a certain layout $\phi^*$ of $G$
% such that
% 
% \begin{gather}
% %
% F(\phi^*, G) = \min_{\phi} F(\phi, G)
% %
% \end{gather}
%
%\subsection{Minimum Linear Arrangement (MLinA)}
%
Now, the definition of Minimum Linear Arrangement problem (MLinA) is as
follows: find a layout $\phi^*$ that minimizes the sum of differences of each
pair of two vertices connected with an edge:
%
%$LA(\phi^*, G) = \min_{\phi} LA(\phi, G)$, or

\vspace{-1em}
\small
\begin{gather}
\overbrace{\sum_{v \in V} \sum_{u \in N_v} |\phi^*(v) - \phi^*(u)|}^{\phi^* \text{minimizes this expression}} = \min_{\forall \phi, \forall v \in V}\ \ \sum_{v \in V} \sum_{u \in N_v} |\phi(v) - \phi(u)| \label{eq:obj_fun_1} 
\end{gather}
\normalsize

A strongly related problem is {Minimum Logarithmic Arrangement}
(MLogA) where ones derives a layout $\phi^*$ that minimizes the sum of
logarithms of differences; incorporating logarithms takes into account the
exact bit count of numbers to be encoded

\vspace{-1em}
\small
\begin{gather}
\overbrace{\sum_{v \in V} \sum_{u \in N_v} \log |\phi^*(v) - \phi^*(u)|}^{\phi^* \text{minimizes this expression}} = \min_{\forall \phi, \forall v \in V}\ \ \sum_{v \in V} \sum_{u \in N_v} \log |\phi(v) - \phi(u)| \label{eq:obj_fun_2} 
\end{gather}
\normalsize

Next, the objective function can also minimize the sum of differences between
consecutive neighbors in adjacency lists (Minimum Linear Gap Arrangement
problem (MLinGapA)), which one can directly use to decrease the storage for a
given graph if differences between vertex ID are stored and encoded with
variable-length coding:

\vspace{-1em}
\small
\begin{gather}
\overbrace{\sum_{v \in V} \sum_{i = 0}^{|N_v|-1} |\phi^*(N_{i+1,v}) - \phi^*(N_{i,v})|}^{\phi^* \text{minimizes this expression}} = \min_{\forall \phi, \forall v \in V}\ \ \sum_{v \in V} \sum_{i = 0}^{|N_v|-1} |\phi(N_{i+1,v}) - \phi(N_{i,v})| \label{eq:obj_fun_3} 
\end{gather}
\normalsize

Finally, the same problem can be (analogously to MLinA) formulated including logarithms and result in 
Minimum Logarithmic Gap Arrangement (MLogGapA):

\vspace{-1em}
\small
\begin{gather}
\overbrace{\sum_{v \in V} \sum_{i = 0}^{|N_v|-1} \log|\phi^*(N_{i+1,v}) - \phi^*(N_{i,v})|}^{\phi^* \text{minimizes this expression}} = \min_{\forall \phi, \forall v \in V}\ \ \sum_{v \in V} \sum_{i = 0}^{|N_v|-1} \log|\phi(N_{i+1,v}) - \phi(N_{i,v})| \label{eq:obj_fun_4} 
\end{gather}
\normalsize

\subsection{Compression Schemes Based on Minimum Arrangement Problems}

There exist many works that reduce the complexity or propose heuristics for the
arrangement problems in~\cref{sec:ilp_defs}. They are listed
in existing surveys~\cite{diaz2002survey, petit2013addenda}; we do not
explicitly describe them as they do not directly relate to graph compression.
Second, various works compress graphs by simply enhancing vertex labelings; we
addressed many of these works in previous sections.
We now only focus on works that explicitly compress graphs using minimum
arrangement ILP formulations. 

Safro and Temkin~\cite{safro2011multiscale} enhance the MLogA for general
graphs by using link weights in the ILP formulation. They motivate
it by observing that link weight can measure how often a link is used;
links that are accessed more frequntly would be compressed more effectively. 
Their algorithm is based on a more generic strategy called the algebraic
multigrid (AMG) methodology~\cite{brandt1982algebraic} for linear ordering
problems~\cite{safro2009multilevel}. In AMG, one first decomposes the original
problem into several approximate ones. In the case of MLogGapA, each
approximate subproblem is based on a projection of the corresponding graph
Laplacian into a lower-dimensional space. Then, solutions of subproblems are
used to derive the final solution. This approach has two key advantages: it
has \emph{``a linear complexity, and it can be relatively easily parallelized and implemented by using
standard matrix–vector operations''}~\cite{safro2009multilevel}. 
Now, Safro and Temkin first formulate the weighted MLogA problem:

\vspace{-1em}
\small
\begin{gather}
\overbrace{\sum_{v \in V} \sum_{u \in N_v} w_{vu} \log |\phi^*(v) - \phi^*(u)|}^{\phi^* \text{minimizes this expression}} = \min_{\forall \phi, \forall v \in V}\ \ \sum_{v \in V} \sum_{u \in N_v} w_{vu} log |\phi(v) - \phi(u)| \label{eq:obj_fun_w} 
\end{gather}
\normalsize

Then, they conduct a series of steps, in each step they reduce the size of the
input graph by \emph{coarsening it}: repeatedly merging pairs of vertices that
satisfy certain properties. At some point, the (much smaller) obtained graph is
used to solve Eq.~(\ref{eq:obj_fun_w}). Then, the original graph is derived by
reversing the coarsening effects, with the computed solution updated at each
de-coarsening step.

Chierichetti et al.~\cite{chierichetti2009compressing} analyze various aspects
of Minimum Arrangement problems; they target social networks but their formal
analysis is generic.
Specifically, they prove that MLogA is NP-hard on multi-graphs (graphs that
admit multiple edges between two vertices), MLinGapA is NP-hard, and that MLogA
has the time lower bound of $\Omega(m \log n)$ for expander-like graphs.
Similarly, Dhulipala et al.~\cite{dhulipala2016compressing} discuss arrangement
problems in the context of graph compression; they offer a proof of the
NP-hardness of MLogGapA and they introduce the ILP formulation of
\emph{Bipartite Minimum Logarithmic Arrangement} (BiMLogA), essentially the
MLogA for bipartite graphs.

Finally, there exist various algorithms that enhance graph compression by
relabeling vertices but without explicitly mentioning the ILP formulation of
arrangement problems~\cite{boldi2011layered, boldi2009permuting_social,
Blandford:2003:CRS:644108.644219}. We covered them
extensively in past sections.

\section{Remaining Schemes}
\label{sec:remaining}

We also discuss schemes that fall outside other categories.
Johnson et al.~\cite{johnson2004compressing} discusses how to compress binary
matrices by reordering the columns so that the whole matrix is more compression-friendly.
Such matrices could be used to represent graphs using less storage.
Moreover, Borici and Thomo~\cite{borici2014semantic} compress graphs by
transforming them into corresponding \emph{hypergraphs} and then partitioning
the hypergraphs so that vertices with similar properties (e.g., degrees) are in
the same partition. This makes the corresponding adjacency matrix more
compression-friendly.
Other works include compressing dense graphs~\cite{kao1998simple} and
vertex-transitive graphs~\cite{litow2004compression}.

\subsection{Hierarchical Schemes}

We discuss general hierarchical schemes similar to those presented
in the web graph section~\cref{sec:hierarchical_web}.

\subsubsection{Grouping Cells of Adjacency Matrix}

First, we outline works that utilize hierarchy related to adjacency
matrices, for example, group non-zero cells into blocks and compress
such blocks separately.
Lim, Kang, and Faloutsos~\cite{kang2011beyond, lim2014slashburn} propose
SlashBurn: a scheme that exploits high-degree vertices (\emph{hubs}, found
often in real-world graphs) and their neighbors (\emph{spokes}) to achieve high
compression ratios. This forms a different type of community structure than the
traditional ``caveman'' communities where vertices are clustered within certain
groups (``caves'') and sparsely connected to other vertex groups. They propose
vertex relabeling that uses this observation and results in space-efficient
representation of the adjacency matrix.
The SlashBurn algorithm (1) removes high-degree vertices and
assign them the lowest labels (2) finds connected components in the resulting
graph and assign the vertices in these components the highest labels, \emph{``in the
decreasing order of sizes of connected components they belong to''}~\cite{lim2014slashburn},
(3) finds the giant connected component in the resulting graph and executes
step~(1) on it recursively, until its size is below a certain threshold.
SlashBurn was extended to distributed-memory settings. \maciej{TODO cite sb-d
Chinese citation}
Moreover, Li and Rao compress graphs by grouping parts of the adjacency matrix and using
different codes to reduce the space required to store a given
group~\cite{li2016compression}.
Furthermore, Li et al.~\cite{li2017efficient} 
first cluster graph adjacency
matrix via graph structure information, and then represent the clustered matrix
by lists of encoded numbers.
Finally, various schemes described in other parts of this survey 
are related to compressing adjacency matrices hierarchically. Examples are
works on $k^2$-trees~\cite{brisaboa2009k2} (see~\cref{sec:k2_web}).

\subsubsection{Schemes Based on Supervertices}

A large portion of hierarchical schemes explicitly groups vertices with similar
properties into \emph{supervertices} (also called \emph{supernodes}) and
collapse edges between them into \emph{superedges}. Many of them were described
in~\cref{sec:hierarchical_web}.  Here, we mention works that are not explicitly
related to web graphs.
Stanley et al.~\cite{stanley2017compressing} find clusters in a given graph and
then simplify and represent it using supervertices with one vertex being formed
from one cluster. 
Next, Toivonen, Zhou, and others~\cite{zhou2012methods,
toivonen2011compression, toivonen2012network} propose a hierarchical scheme that targets weighted
graphs. They group vertices with similar neighborhoods into supervertices, and
group multiple weighted edges between such supervertices into superedges. 
Another similar work that considers algorithms for bipartite matching, edge connectivity, vertex connectivity, and all-pairs shortest paths,
was conducted by Feder and
Motwani~\cite{feder1991clique, feder1995clique}. 
Moreover, Brown et al.~\cite{brown2016evolving} use genetic algorithms to
assess the similarity of vertices (where two vertices are considered similar if
many of their neighbors are identical). Similar vertices are merged into
supervertices and the graph size is ultimately reduced.
In addition, Sun et al.~\cite{sun2008graph} measure the overlap of neighbors
between vertices and, if the overlap is large enough, the identical
neighborhood parts are collapsed and a certain data structure is used to encode
this structural change.
Lamarche-Perrin et al.~\cite{lamarche2016information}
target compressing weighted graphs with supervertices.
Finally, Nourbakhsh simplifies the input graph (and thus reduces its size). He
uses Szemeredi's Regularity Lemma~\cite{komlos1996szemeredi} to cluster the graph and to
produce a smaller graph where clusters become
vertices~\cite{nourbakhsh2015algorithms}.

\subsubsection{Tree Decompositions}

Some lossless compression schemes decompose a graph into several trees, encode
these trees separately, and ultimately reduce the overall space requirements.
Chen and Reif~\cite{Chen:1996:ELC:789084.789454} decompose an input graph into
several binary trees, and finally compress these trees with a proposed
tree-compression algorithm.
The key idea in compressing a single binary tree is to further decompose this
tree into smaller subtrees. These subtrees are small enough that any such
subtree can be found multiple times in the input tree. Thus, after the full
binary tree decomposition, the authors calculate occurrence probabilities for
each subtree and assign the corresponding Huffman code to it.  Finally, the
tree is encoded by traversing it and assigning the above codes.
Now, the method to find and count respective subtrees is similar to counting
words in texts. Specifically, the authors traverse the input tree with BFS and
build a suffix tree in the process where each node of a suffix tree corresponds
to one specific subtree.
A similar approach for compressing probabilistic graphs was described by Maniu
et al.~\cite{maniu2013compression}.

\subsubsection{Others}

Feder et al.~\cite{feder2003representing}
proposed approximation algorithms for obtaining the best virtual-node hierarchical
compression.
They
also illustrated that the optimal compression of this type is NP-hard.
Other works include compression used for obtaining better
clustering~\cite{mueller2011weighted, cilibrasi2005clustering}, using
quadtrees to compress adjacency matrices~\cite{chatterjee2016exploiting},
compressing graphs that model automata~\cite{mohri2015automata},
partitioning an input graph and compressing each partition independently~\cite{dhabu2013partition}.

\subsection{Compression for More Efficient Computation}

Here, we outline works that specifically use compression for faster graph
algorithms. These schemes are different from the ones described in the section
devoted to problem-aware graph compression (\cref{sec:problem-aware}) because
they do not propose novel compression but discuss how to use existing
compression schemes for faster graph algorothms.
Liakos et al.~\cite{liakos2016memory} use various compression techniques (bit
vectors and different types of coding techniques) in distributed-memory graph
processing engines to reduce the pressure on the memory subsystem and thus
accelerate processing.
Next, Granskog and Striger analyze whether graph traversal algorithms (BFS, DFS) can
be accelerated by using compression methods such as
$k^2$-trees~\cite{granskog2015comparison}.
Other works use compression as one of the tools for better data mining
capabilities~\cite{feng2013compression}, faster queries on
graphs~\cite{nabti2017querying}, accelerating subgraph matching by reducing
the size of sets that contain matching candidates~\cite{qiao2017subgraph},
or solving bin packing problems more efficiently~\cite{brandao2016bin}.

\subsubsection{Compression in Graph Processing Engines}

Some works specifically discuss how to accelerate a given graph processing
engine with compression.
Shun et al.~\cite{shun2015smaller, shun2015smaller} developed Ligra+, an
enhancement over the Ligra graph processing engine~\cite{shun2013ligra} that
uses parallelism to accelerate compression and decompression of graph data and
thus amortize the costs of utilizing compressed graph representations while
reducing the pressure on the memory subsystem.
Other works that use parallelization to accelerate compression also
exist~\cite{en2018parallel}.
Furthermore, Chen et al.~\cite{chen2008graph} used compressed graphs with a
generic topological OLAP framework in online graph analysis.
Another paper~\cite{aberger2015emptyheaded} uses various vertex relabelings for
more compression friendly graph layout within EmptyHeaded, an engine that
outperforms standard OLAP systems.
Chavan conducted an empirical study on graph compression in engines such as Pregel~\cite{Malewicz:2010:PSL:1807167.1807184, chavanintroduction}.

\subsection{Vertex Coding}

Certain papers from 60s are tentatively connected to graph compression.
Specifically, Breuer and Folkman~\cite{breuer1967unexpected, breuer1966coding}
analyzed \emph{coding vertices}, i.e., assigning each vertex a unique binary
code that is always smaller than a certain constant if two vertices are
connected, and is always larger than this constant if two vertices are not
connected.
Using these coding schemes, 
one can determine the adjacency of any two vertices by using
the Hamming distance of their labels. This may have applications in the domain of
implicit graph representations (\cref{sec:implicit}).

\section{Related Domains Covered in Surveys}
\label{sec:others_surveyed}

We now mention works and surveys covering areas that are related to lossless
graph compression.
First, there are works on compressing graphs with the purpose of \macb{more
effective visualization}, for example Dwyer et al.'s~\cite{dwyer2013edge}.
They were partially covered in another survey~\cite{liu2016graph}.
Second, \macb{compression of meshes} was covered extensively in several
surveys~\cite{maglo20153d, peng2005technologies, szymczak2002piecewise, aleardicompact}.
Third, \macb{compression of trees} is outside the scope of this work. It is
partially covered in other works~\cite{katajainen1990tree}.

\subsection{Lossless Summarization of Graphs}
\label{sec:summarization_of_g}

\emph{Summarization of graphs} is an area where graph is
\emph{summarized} to provide a smaller graph description that may
focus on some particular graph aspects~\cite{shah2015timecrunch, khan2015set,
khan2014set, khan2017set, khan2017faster, liureducing, rivas2014managing,
ashrafi2017graph}. These works were covered in a survey~\cite{liu2016graph}.
The most important connection to graph compression is that in many of these
schemes, the process of graph summarization also leads to size reduction. For
example, vertices within a cluster are grouped to form a supervertex, and edges
are merged into superedges~\cite{wu2014graph, liu2014distributed,
tian2008efficient, bei2016summarizing, liu2017graph, riondato2017graph,
payaman2017gssc, zhang2010discovery, you2013towards, zhou2017summarisation,
koutra2014vog, koutra2015summarizing, lefevre2010grass}, similarly to many
hierarchical schemes in web graphs (\cref{sec:hierarchical_web}). Some works
use or discuss bisimulation, especially in the domain of RDF
graphs~\cite{vcebiric2015query, campinas2016graph}. In addition, various works
use summarization to better understand the structure of graphs in domains such
as biology~\cite{navlakha2009revealing, toivonen2010framework} or independently
of a specific domain~\cite{ahnert2014generalised}.
Others focus on illustrating the impact of simplification on graph
structure~\cite{blagus2014assessing, blaguslarge}.
Moreover, there are works dedicated to the \macb{summarization of dynamic
graphs}~\cite{khan2017summarizing_d, tsalouchidou2016scalable,
qu2016efficient}.

\subsection{Efficient and Implicit Graph Representations}
\label{sec:implicit}

Intuitively, \emph{efficient} (in many cases also called \emph{implicit}) graph
representations provide vertex labels that encode the structure of the input
graph so that no additional structure dedicated to storing edges is required.
For example, Kannan and Naor in their seminal work~\cite{kannan1992implicat}
assign $\mathcal{O}(\log n)$ bit labels to vertices such that these labels
completely encode the structure of the graph. Thus, no additional data
structure that determines edges is required.  In addition, given the labels of
any two vertices, the authors show that one can test if the vertices are
adjacent in time linear in the size of the labels. 
Many other such schemes exist~\cite{peleg2000informative, talamo1998compact,
alstrup2002small, scheinerman1999local}.
Another thread of related work are algorithms for efficient derivation of such
representations~\cite{arikati1997efficient}.
Now, such representations are covered extensively in a book by
Spinrad~\cite{spinrad2003efficient}.
Since the book was published, more such representations were
discovered~\cite{gavoille2007shorter, chandoo2016implicit,
talamo2001representing, curtis2010implicit}.

\macb{Terminology Clarification}
We now clarify a certain terminology issue.  An \emph{implicit} graph
representation as described above is a representation where vertex labels
themselves \emph{encode the information on edges between vertices}.
Now, the term \emph{implicit} is used in another context in the
literature~\cite{succ-category}. It describes a representation of an arbitrary
data that is a constant additive factor away from the storage lower bound for
this data. Formally, if the optimum to store some data is $\mathcal{N}$ bits,
an implicit representation takes $\mathcal{N} + \mathcal{O}(1)$
bits~\cite{succ-category}.
We do not know of any graph representations that are implicit in the second
sense.

\section{Taxonomy and Discussion of Features}
\label{sec:features_taxonomy}

We now group and discuss graph compression schemes based
on selected common \emph{features} for better understanding of lossless graph
compression.

\subsection{Problem-Aware Graph Compression}
\label{sec:problem-aware}

First, we discuss schemes that, despite applying compression, still allow to 
obtain selected graph properties or solve selected graph problems fast.
Sadri et al.~\cite{sadri2017shrink} propose Shrink, a compression scheme that
preserves distances between vertices. The compression proceeds in steps, in
each step it iteratively merges vertices. During each merging, a system of
linear equations is solved to define new edge weights to minimize changes in
the distances. Merging continues until a specified number of vertices is
reached. 
Moreover, Fan et al.~\cite{fan2012query, fan2012graph} develop compression
strategies that preserve high performance and losslessness for two classes of
graph queries: reachability and graph pattern queries via (bounded) simulation.
Next, Hernandez discusses application-driven graph representations and
compression~\cite{hernandez2009managing}.
Another similar work (performed for general graphs) that considers algorithms
for edge connectivity, vertex connectivity, all-pairs shortest paths, and bipartite matching, 
was conducted by Feder and Motwani~\cite{feder1991clique,
feder1995clique}. 
Finally, most of succinct and compact graph representations provide graph
queries that ensure a specific time complexity, most often constant-time or
logarithmic, see~\cref{sec:bounds}.

\subsection{Compression of Dynamic Graphs}
\label{sec:dynamic}

We separately discuss compressing \emph{dynamic} graphs (in the literature,
they are also called \emph{temporal}, \emph{evolving}, or
\emph{time-evolving}).
% 
% There are only a few such papers as the vast majority of schemes do
% not consider graph updates.
%
Brodal and Fagerberg~\cite{brodal1999dynamic} present a linear space graph data
structure for graphs with bounded arboricity (example such graphs are planar
graphs or bounded-treewidth graphs) under insertions, edge deletions, and
adjacency queries. 
The proposed representation is the adjacency list representation,
augmented with a simple scheme that maintains the structure under graph
modifications. The core idea in proving the stated time bounds (constant-time
adjacency queries in a graph with bounded arboricity $c$) is to reduce this
problem to a simpler problem of assigning directions to edges (i.e.,
constructing a directed graph out of the input undirected one) so that all
vertices have outdegree $\mathcal{O}(c)$.

Iverson and Karypis~\cite{iverson2014storing} propose five data structures for
representing dynamic sparse graphs. Their structures offer different trade-offs
between size and speed of provided graph operations.  The structures are:
Linked-List (LL), Batch Compressed Sparse Row (BCSR), Dynamic Adjacency Array
(DAA), Dynamic Intervalized Adjacency Array (DIAA), and Dynamic Compressed
Adjacency Array (DCAA).  LL is based on a simple set of linked lists with one
list being responsible for one vertex neighborhood.  BCSR is essentially an LL, but
when a size of a linked list grows too large, it is resized into a static CSR.
In DAA, each neighborhood is a dynamically allocated array that must be resized
if there are updates.  DIAA is essentially a DAA enhanced with storing
contiguous vertices as intervals.  Finally, DCAA leverages ideas from the
WebGraph framework (\cref{sec:web_graphs_web_graph}).

Other works on dynamic graphs exist, for example Boldi et
al.~\cite{boldi2008large} analyzes how the web graph evolves and how its
respective snapshots can be compressed, Caro et al.~\cite{caro2015data}
design compact graph representations that enable answering queries fast,
and Klitzke and Nicholson~\cite{klitzke2016general} mention
compressing dynamic graphs as a part of their general framework for managing
dynamic succinct data structures.

Finally, several dynamic schemes were described in the other sections of this
survey, for example in the parts devoted to summarizing dynamic
graphs~\cite{khan2017summarizing_d, tsalouchidou2016scalable, qu2016efficient,
shah2015timecrunch} (\cref{sec:summarization_of_g}), compressing RDF graphs
(\cref{sec:rdf_graphs}), De Bruijn graphs~\cite{belazzougui2016fully}
(\cref{sec:de_bruijn_biology}), graph databases~\cite{labouseur2015g}
(\cref{sec:databases}), succinct data structures~\cite{tamassia1988dynamic}
(\cref{sec:bounds}), and others~\cite{brisaboa2017compressed,
schiller2015efficient, mondal2015real}.  We list them to facilitate
navigating the survey and refer the reader to these specific sections for more
information.

Dynamic graphs are also considered in streaming settings; we discuss this
separately in~\cref{sec:streaming_feature}.

\subsubsection{Viewing Graphs As Tensors}

We separately discuss works that add more dimensions to its adjacency matrix to
model changes.
Caro et al.~\cite{caro2016compressed} represent dynamic (\emph{temporal})
graphs using 4-dimensional binary tensors. Two dimensions are used to model
edges and two other dimensions model time intervals where a given edge exists.
Then, they propose to compress such a representation with a generalization of
$k^2$-trees~\cite{brisaboa2009k2} (see~\cref{sec:k2_web}) to a $d$-dimensional
space, called $k^d$-tree.  The key idea is similar to that of simple
$k^2$-trees, namely, parts of a $d$-dimensional tensor with zeros in cells are
compressed with internal nodes of a $k^d$-tree while tree leaves represent
parts of the tensor that have more than one non-zero cell.
Related approaches based on viewing a dynamic graph as a 4-dimensional object
were discussed by Brisaboa, Bernardo, Caro, and others~\cite{de2013compact,
brisaboa2014compressed_t}.

\subsection{Compression of Graphs in Streaming Settings}
\label{sec:streaming_feature}

Various compression schemes are designed for streaming
settings~\cite{packer2017graphzip, tang2015summarizing, sekharan2017queryable,
khan2017toward, zhang2016compressing, nelson2015compressing, tang2016graph,
khan2016query, demetrescu2009trading, lin2011scent}.  For example, Nelson et
al.~\cite{nelson2015compressing} use quadtrees to compress graph streams.

\section{Conclusion}

Graph compression is an important area of research as it can be used to
accelerate numerous modern graph workloads by reducing the amount of
transferred data. Yet, it is a diverse set of fields driven by different
communities, with a plethora of techniques, algorithms, domains, and
approaches.
We present the first survey that analyzes the rich world of
lossless graph compression.  We do not only list and categorize the existing
work, but also provide key ideas, insights, and discuss formal underpinning of
selected works.
Our work can be used by architects and developers willing to select the best
compression scheme in a given setting, graph theoreticians aiming to understand
the high-level view of lossless graph compression, and anyone who wants to
deepen their knowledge of this fascinating field.  

\vspace{0.5em}
\noindent
\macb{ACKNOWLEDGEMENTS }
We thank Olivier Devillers, Hsueh-I Lu, Miguel A. Martínez Prieto, Luca Castelli Aleardi, Gonzalo Navarro,
and Sebastiano Vigna for their insightful comments.

\renewcommand*{\bibfont}{\small}

% Bibliography
{\sf
%\scriptsize
%\bibliographystyle{ACM-Reference-Format}
\bibliographystyle{abbrv}
\bibliography{refs}
%\bibliography{../refs_ultrashort}
}

\end{document}